\documentclass[iop,apj,twoside]{emulateapj}

\usepackage{xspace}
\usepackage{amssymb}
\usepackage{times}
\usepackage{verbatim}
\usepackage{color}

\newcommand{\mone}  {^{-1}}
\newcommand{\mtwo}  {^{-2}}
\newcommand{\mthree}{^{-3}}

\newcommand{\pthree}{^3}

\newcommand{\cmmtwo}{\,\mathrm{cm\mtwo}}

\newcommand{\cmmthree}{\,\mathrm{cm\mthree}}
\newcommand{\cmpthree}{\,\mathrm{cm\pthree}}

\newcommand{\cps}   {\,\mathrm{counts\,s\mone}}

\newcommand{\kev}   {\,\mathrm{keV}}
\newcommand{\ev}    {\,\mathrm{eV}}

\newcommand{\ks}    {\,\mathrm{ks}}
\newcommand{\pc}    {\,\mathrm{pc}}
\newcommand{\lum}   {\,\mathrm{ergs\,s\mone}}
\newcommand{\mang}  {\,\mathrm{{\AA}}\xspace}
\newcommand{\mmang}  {\,\mathrm{{m\AA}}\xspace}
\newcommand{\rmang} {\mathrm{{\AA}}\xspace}

\newcommand{\mk}    {\,\mathrm{MK}}

\newcommand{\apflux} {\,\mathrm{phot\,cm\mtwo\,s\mone}}

\newcommand{\arf}   {{ARF}\xspace}

\newcommand{\chan}  {{\it Chandra}\xspace}

\newcommand{\ciao}  {{CIAO}\xspace}

\newcommand{\heg}   {{HEG}\xspace}
\newcommand{\hetgs} {{HETGS}\xspace}
\newcommand{\hetg}  {{HETG}\xspace}

\newcommand{\meg}   {{MEG}\xspace}

\newcommand{\rgs}   {{RGS}\xspace}
\newcommand{\rmf}   {{RMF}\xspace}

\newcommand{\xmm}   {{\it XMM-Newton}\xspace}
\newcommand{\hrtennn} {{HR$\,1099$}\xspace}
\newcommand{\hrten}   {\hrtennn}
\newcommand{\hrtenn}  {\hrtennn}
\newcommand{\siggem} {{$\sigma\,$Gem}\xspace}

\newcounter{ion} \newcommand{\eli}[2]{\setcounter{ion}{#2}#1{~\sc\roman{ion}}}

\shorttitle{Solar and Stellar Flare X-ray Spectra}
\shortauthors{Huenemoerder et al.}

\begin{document}

\title{Stellar Coronae, Solar Flares: a Detailed Comparison of
  \siggem, \hrtennn, and the Sun in High-resolution X-rays}

\author{David P.\ Huenemoerder\altaffilmark{a},
Kenneth J.\ H.\ Phillips\altaffilmark{b},
Janusz Sylwester\altaffilmark{c},
Barbara Sylwester\altaffilmark{c}
\ \\
}

\altaffiltext{a} {Massachusetts Institute of Technology, Kavli
  Institute for Astrophysics and Space Research, 70 Vassar St.,
  Cambridge, MA, 02139, USA (\url{dph@space.mit.edu})}

\altaffiltext{b} {Visiting Scientist, Space Research Center, Polish
  Academy of Sciences, 51-622, Kopernika~11, Wroc{\l}aw, Poland
  (\url{kennethjhphillips@yahoo.com}) }

\altaffiltext{c} {Space Research Center, Polish Academy of Sciences,
  51-622, Kopernika~11, Wroc{\l}aw, Poland
  (\url{js@cbk.pan.wroc.pl}, \url{bs@cbk.pan.wroc.pl})}


\begin{abstract}
%
{ 
    The \chan High Energy Transmission Grating Spectrometer (HETG)
    spectra of the coronally active binary stars \siggem and \hrtennn
    are among the highest fluence observations for such systems taken
    at high spectral resolution in x-rays with this instrument.  
} 
This allows us to compare their properties in detail to solar flare
spectra obtained with the Russian {\it CORONAS-F} spacecraft's RESIK
instrument at similar resolution in an overlapping bandpass.  Here we
emphasize the detailed comparisons of the $3.3$--$6.1\mang$ region
(including emission from highly ionized S, Si, Ar, and K) from solar
flare spectra to the corresponding \siggem and \hrtennn spectra.  We
also model the the larger wavelength range of the HETG, from
$1.7$--$25\mang$---having emission lines from Fe, Ca, Ar, Si, Al, Mg,
Ne, O, and N---to determine coronal temperatures and abundances.
\siggem is a single-lined coronally active long-period binary which
has a very hot corona. \hrtennn is a similar, but shorter period,
double-lined system.  With very deep HETG exposures we can even study
emission from some of the weaker species, such as K, Na, and Al,
which are important since they have the lowest first ionization
potentials, a parameter well known to be correlated with elemental
fractionation in the solar corona.  The solar flare temperatures reach
$\approx 20\mk$, comparable to the \siggem and \hrtennn coronae.
During the Chandra exposures, $\sigma\,$Gem was slowly decaying from a
flare and its spectrum is well characterized by a collisional
ionization equilibrium plasma with a broad temperature distribution
ranging from 2--60 MK, peaking near 25 MK, but with substantial
emission from 50 MK plasma.  We have detected \eli{K}{18} and
\eli{Na}{11} emission which allow us to set limits on their
abundances.  \hrtennn was also quite variable in x-rays, also in a
flare state, but had no detectable \eli{K}{18}.  These measurements
provide new comparisons of solar and stellar coronal abundances,
especially at the lowest FIP values.  The low FIP elements do not show
enhancement in the stellar coronae as they do in the Sun, except
perhaps for K in \siggem.  While \siggem and \hrtennn differ in their
emission measure distributions, they have very similar elemental
abundances.
\end{abstract}




\section{Introduction}

Accurate knowledge of elemental abundances is of fundamental
importance for many astrophysical processes.  In soft x-rays from hot,
low-density plasmas ($\lambda \sim1$--$100\mang$, or
$E\sim0.1$--$12\kev$; $T \sim 1$--$100\mk$; $n_e\lesssim
10^{12}\cmmthree$) in collisional ionization equilibrium,
emission processes are relatively simple in that every collisional
excitation results in a radiative transition whose photon escapes the
plasma.  At coronal temperatures, ions are highly charged and the
dominant species are typically H- and He-like ions.  The abundant
elements from C to Fe have their principal H- and He-like lines in
this band.  Hence, high-resolution x-ray spectra provide a wealth of
emission lines from numerous abundant elements over a range of
temperatures from a plasma which is relatively easily modeled.

The solar corona and coronae of other stars still have some
unexplained properties related to their elemental abundances, and
these, in turn, relate to the formation of coronae and their use as
diagnostics of stellar abundances.  It has been known for quite some
time that there is a correlation of solar coronal abundances with the
elements' first ionization potentials (FIP) in the sense that elements
with low FIP ($\lesssim 10\ev$) are enhanced by up to a factor of 4
over photospheric values \citep{Feldman:92, Feldman:Laming:00}.

In other stars, coronal abundance anomalies were suspected even with
low-resolution {\it ASCA} spectra \citep{Kaastra:96}, but there the
low-FIP elements were depleted relative to solar photospheric values
(termed an inverse FIP effect).  With the advent of high-resolution
spectroscopy with the \chan X-ray Observatory (CXO) and \xmm grating
instruments nearly a decade ago, these results were confirmed in
detail \citep{Audard:01a, Brinkman:01}.  The stellar observations have
several difficulties in interpretation and in direct comparison to the
Sun.  The photospheric abundances are often unknown, so the comparison
is done against solar values.  The stellar disks are unresolved; we do
not know whether particular structures, such as actively flaring
regions or quiescent loops, dominate the emission, nor how extended
are such structures.  Since flares in the Sun---and presumably other
stars---heat and evaporate chromospheric plasmas, it has been
suggested that the flare matter would more likely represent the
underlying photospheric plasma, since purported diffusion processes
sensitive to the FIP have had no time to have effect.  Observational
results have been mixed. \citet{Nordon:Behar:2008} found no strong or
consistent correlation of abundance changes during stellar flares.
{From low-resolution, few-temperature-component models,
  \citet{Liefke:al:2010} found a factor of three increase in Fe
  abundance over the quiescent value during a large flare on CN~Leo;
  they could detect no effect in other elements.}
\citet{Wood:Linsky:2010} and \citet{Wood:al:2012} examined the FIP
effect in M-dwarfs and found that both high and low activity stars can
have an inverse FIP effect.  They suggested that there is a trend
with spectral type and that stars of type K are unbiased, earlier
types have a FIP effect, and later types have an inverse FIP effect.

\citet{Laming:Hwang:2009} and \citet{Laming:2012} have presented a
theoretical basis for low-FIP ion fractionation in coronal loops via
ponderomotive forces associated with Alfv\'en waves passing through the
loops.  Some elements can be either enhanced or depleted in the corona
relative to the photosphere, depending upon the direction of the
ponderomotive force.  Calculations of fractionation amounts with model
loops having typical parameters are similar to those observed.

The determination of abundances from x-ray spectra can have
far-reaching implications.  \citet{Drake:Testa:2005} used the rather
uniform values of the Ne:O abundance ratio in many stars to argue
that these elements do not undergo differential fractionation, but
represent the cosmic abundance ratio, and so adoption of this ratio
for the Sun would reconcile a rather serious conflict with stellar
interior models and helioseismology.

Here we present new results for the lowest FIP elements, K and Na, in
the \chan/\hetg spectra of \siggem and \hrtennn.  By modeling the
\hetgs spectra over the $1$--$25\mang$ range, we also derive the
emission measure distributions and elemental abundances of all the
major contributors, which span a broad range in FIP.  These models
improve on prior work on these \chan spectra \citep[e.g.][]{Drake:01,
  Nordon:Behar:2007, Nordon:Behar:2008}.

To compare \siggem and \hrtenn to the Sun, we use a high-resolution
x-ray spectrum from the {\it CORONAS-F} RESIK instrument of a solar
flare on 2002 December 26, in which the high-temperature plasmas
reach the mid-range of temperatures found in the stellar spectra
whose emissions are also probably due, in large part, to stellar
flares, 
%
{ 
  as we will show for \siggem here, and as has been shown for \hrten
  by \citet{Nordon:Behar:2007}.  } 
%

\section{Observations and Calibration}

\subsection{\chan/\hetg}

The \chan/\hetg instrument \citep{HETG:2005} observed \siggem and
\hrtennn twice each; dataset identifiers and exposure times for each
star are given in Table~\ref{tbl:stars}.  The \hetgs spectra cover the
range from about $1$--$30\mang$, as dispersed by two types of
grating facets, the High Energy Grating (\heg) and the Medium Energy
Grating (\meg), with resolving powers of 
{between about 100 to 1000}, with
approximately constant full-width-half-maxima (FWHM) of $12\mmang$
for \heg and $23\mmang$ for \meg.  
{These two objects have the highest fluence exposures among
  hot, coronally active stars observed with the \hetgs; there are
  about 350,000 counts for each star in the combined first orders over
  the $2$--$15\mang$ range covered by both \meg and \heg.}
The only other similar source with more counts is Capella (a
calibration object) having 20 \hetgs exposures for nearly 800,000
counts.  Capella, however, is much cooler than \siggem and \hrtennn
and has relatively little flux in the $3$--$6\mang$ region---about
10\% that of the other two stars.

%
\begin{deluxetable}{c c c}
  \tablecolumns{3}
  \tablewidth{0.95\columnwidth}
  \tablecaption{Stellar/Observational Information\label{tbl:stars}}
  \tabletypesize{\small}
  \tablehead{
    \colhead{Property}&
    \colhead{\siggem}&
    \colhead{\hrtennn}
  }
  \startdata
  \chan dataset IDs& 5422, 6282& 1252, 62538\\
  Date Obs& 2005-05-16,17& 1999-09-14,17 \\
  \hetg Exposures [ks]&   62.8, 57.9&  14.7, 94.7\\
  Spectral Type& K1 III + ? & K1 IV + G5 IV\\
  $d [\pc]$& 37.5& 30.68\\
  $N_\mathrm{H}\,[10^{18}\cmmtwo]$& 1.0& 0.94\\
  $L_{bol} [10^{34}\lum]$& $21.4$& $3.23$\\
  $L_x [10^{31}\lum]$& $3.0$& $1.9$\\
  $VEM\,[10^{54}\cmmthree]$&  2.4& 1.5
%
  \enddata
  \tablecomments{Observation dates are the UT day on which the
    observation started.
    %
{ 
      Spectral types were taken from
      the compilation of \citet{CABS3}. The neutral hydrogen column
      densities are from \citet{Sanz-Forcada;Brickhouse:02}.
    } 
    Distances are from HIPPARCOS \citep{vanLeeuwen:2007}. Bolometric
    luminosities and volume emission measure ($VEM$) were derived
    using information in \citet{Strassmeier:2009}.  X-ray luminosities
    are from the model spectra derived herein over the $1$--$50\mang$
    ($0.2$--$12\kev$) band.}
\end{deluxetable}

The \chan data were reprocessed with standard Chandra Interactive
Analysis of Observations (\ciao) programs
\citep{CIAO:2006} to apply the most recent calibration data (\ciao 4.4
and the corresponding calibration database as of 2012 June 19).  Since
these are very bright sources with saturated and distorted zeroth
order images, and since the zeroth order provides the origin for the
wavelength scale, we determined the zeroth order centroid from the
intersection of a grating spectrum with the strong zeroth order CCD
frame-shift streak.  The default binning was adopted, which
over-samples the instrumental resolution by about a factor of 4.  The
counts spectra are thus composed of 4 orders per source per
observation: the $\pm1$ orders for each grating type, the \meg and
\heg, which have different efficiencies and resolving powers.

Several calibration files are required for analysis to convolve a
model flux spectrum with the instrumental response in order to produce
model counts.  These are made for each observation and each spectral
order by the \ciao programs which use observation-specific data in
conjunction with the calibration files to make the effective area
files (``Auxiliary Response File'', or \arf) and the spectral
redistribution and extraction-aperture efficiency files (``Response
Matrix File'', or \rmf) \citep{DavisJE:2001b}.

\begin{figure}[!htb]
  \centering\leavevmode
  \includegraphics[width=1.20\columnwidth, angle=-90, viewport=20 20 720 600]{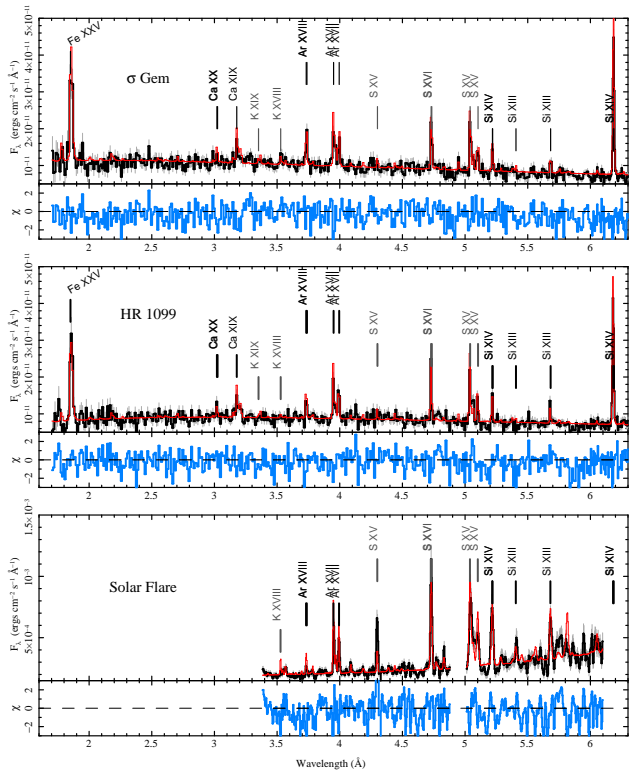}
  \caption{Here we show the spectra and models in the short wavelength
    region where \hetg and RESIK spectra overlap.  Flux-corrected
    spectra are in black, and the red is the model convolved by the
    instrumental resolution.  Below each, in blue, are residuals.  The
    top panel shows \siggem; the middle is \hrtennn; the bottom is the
    rise phase of solar flare on 2002 December 26 (maximum at 06:30
    UT).  All models were evaluated using AtomDB emissivities, though
    the solar spectrum was fit using CHIANTI
    \citep{Dere:Landi:al:2009}.  Residuals shown for the solar
    spectrum include an arbitrary scale factor.  Line identifications
    for prominent or important ions are given.  Emission measures and
    abundances used in the models are given in Figure~\ref{fig:emd}
    and Table~\ref{tbl:abund}, respectively.}
  \label{fig:threespec}
\end{figure}

We show a portion of the \siggem and \hrtennn spectra in
Figure~\ref{fig:threespec} (top and middle panels).

%


\subsection{RESIK}

RESIK (REntgenovsky Spekrometr s Izognutymi Kristalami) was a bent
crystal spectrometer on the Russian {\it CORONAS-F} spacecraft viewing
solar active regions and flares, and was operational between 2001 and
2003.  The instrument consisted of a pair of Si (crystal plane 111)
crystals and a pair of quartz ($10\bar10$) crystals, with a
one-dimensional position-sensitive proportional counter for each
crystal pair.  The combination of bent crystals and position-sensitive
detector enables the entire spectral range to be observed
simultaneously within a data-gathering interval. The duration of these
intervals varied inversely with the amount of incident x-ray emission,
typical values being $2\,$s for flare peaks and up to five minutes for
times late in a flare decay. The wavelength ranges of the four
channels for on-axis sources were 3.40--$3.80\mang$ (channel 1),
3.83--$4.27\mang$ (2), 4.35--$4.86\mang$ (3), and 5.00--$6.05\mang$
(4).  The lack of a collimator allowed RESIK to observe off-axis flares,
extending the wavelength limits in one direction or the other by up to
$70\mmang$. The low atomic number of the crystal material ensured a
much lower background due to crystal fluorescence (which depends on
$Z^4$) than previous solar crystal spectrometers, enabling the solar
continuum to be measured for at least the two shorter-wavelength
channels.  The total wavelength range (3.4--$6.05\mang$) included
resonance lines of He-like K (\eli{K}{18}), H-like and He-like Ar
(\eli{Ar}{18}, \eli{Ar}{17}), He-like Cl (\eli{Cl}{16}), H-like and
He-like S (\eli{S}{16}, \eli{S}{15}), and H-like and He-like Si
(\eli{Si}{14}, \eli{Si}{13}). The wavelength resolution (FWHM) varied
from $8\,\mathrm{m\AA}$ (near $3.3\mang$) to $17\,\mathrm{m\AA}$ (near
$6\mang$). 
{RESIK was unable to resolve thermal Doppler broadening for
  typical flare temperatures. Thus, for a temperature of $15\mk$ the
  thermal Doppler broadening (FWHM) of the \eli{K}{18} resonance line
  at $3.53\mang$ is $1.6\,\mathrm{m\AA}$, and for the \eli{Si}{14}
  line at $5.22\mang$ it is $2.7\,\mathrm{m\AA}$. Further instrumental
  details about RESIK are given by \cite{Sylwester:al:2005}.}

{The RESIK spectra thus cover a much more limited wavelength
  region than the Chandra/HETG spectra but the high quality of the
  absolute calibration allow a detailed comparison of the abundances
  of Si, S, Ar, and K in particular. 
}

In the bottom panel of Figure~\ref{fig:threespec} we show a RESIK
spectrum taken over a data-gathering interval of $86\,$s during the
rise phase of a GOES class~C1.9 flare on 2002 December 26 at 06:30~UT
(maximum).  Identifications of some of the principal emission lines
are given.

\section{Analysis}

\subsection{\hetg Spectral Fitting and Modeling}\label{sec:hetgspecfit}

The analysis of the \hetg spectra required an iterative combination of
global plasma model fits, parametric line fitting, and line-based
reconstruction of the emission measure distribution (EMD) and
elemental abundances.  The first step was to fit a plasma model to the
binned spectrum in order to provide a physically-based continuum for
parametric line fitting.  Even at \heg resolution ($FWHM =
12\mmang$), there are regions of the spectrum for which the apparent
continuum is above the true value due to line blending.  The continuum
originates from free-free (FF) and free-bound (FB) components, and the
latter depends significantly upon elemental abundances, especially for
solar flares \citep{Phillips:Sylwester:al:2010}; for the stars with
high temperatures and low metallicity, FF emission dominates the
continuum.  

{The line emissivities, more generally called $G(T_e)$
  functions (in which $T_e $ is the plasma's electron temperature),
  are defined by the amount of radiation emitted by an optically thin
  plasma per second with unit volume emission measure ($ N_e
  N_\mathrm{H} V$ where $N_e$ is the electron density, $N_\mathrm{H}$
  the proton density, and $V$ the emitting volume). In terms of plasma
  and atomic quantities, we can write $G(T_e)$ for a particular line
  of element, $X$, as}
%
\begin{equation}\label{eqn:gt}
\small
  G(T_e) =  \frac{ N(X^{+n}_{i}) } { N(X^{+n}) }
            \frac{ N(X^{+n}) }  { N(X) }
            \frac{ N(X) } { N(\mathrm{H}) }  
            \frac{ A_{i1} } { N_e }\,
                     \quad\mathrm{[phot\cmpthree s\mone]}
\end{equation}
%
{where $N(X^{+n}_i)$ is the population of the
  excited level-$i$ of an $n$-times-ionized atom, $X^{+n}_i$,
  $N(X^{+n}) / {N(X) }$ is the ion fraction as a function of
  electron temperature from ionization balance calculations, $N(X) /
  N(\mathrm{H})$ the abundance of element $X$ relative to hydrogen, 
  and $A_{i1}$ the transition probability from level $i$ to the ground
  state.}

{The line luminosity is related to $G(T)$ through an integral
  over the emission measure, which, through a change of variables, is
  expressed as a differential in electron temperature instead of
  volume:
}
\begin{equation}\label{eqn:emd}
\small
  L(X_{i}^{+n}) = \int{\! G(T_e)\,
                [ N_e N_\mathrm{H} \frac{\mathrm{d}V} {\mathrm{d}\log T_e} ]\,
                 \mathrm{d}\log T_e }
                     \quad\mathrm{[phot\,s\mone]}
\end{equation}
%
{
(where $i$ and $n$ identify the transition and ion, as in the
expression for $G(T_e)$).  The quantity in square brackets is called
the differential emission measure.  When we refer to the emission
measure distribution (EMD) it will be this quantity integrated over
intervals of $\log T_e$ {\it vs.}\ temperature \citep[for historical
examples, see][]{Lemen:1989, Jordan:1975, Pottasch:1963}.
}

For the plasma models 
{describing the \hetg spectra}, we relied
on the output of the Astrophysical Plasma Emission Code (APEC), as
available in the atomic database, AtomDB \citep{Smith:01,
  Foster:Smith:Brickhouse:2012}, which provided continuum and line
emissivities for low-density plasmas in collisional ionization
equilibrium.  We adopted the defaults for ionization balance
\citep{Mazzotta:98} and solar photospheric abundances
\citep{Anders:89} (though abundances were free parameters and
referenced to more recent determinations later).  The initial fit used
a broken powerlaw 
{model for the emission measure temperature
  distribution} with variable abundances for the most abundant
elements with strong emission line features in the \hetg spectrum
(e.g., N, O, Ne, Mg, Si, S, and Fe).  This model allowed a range of
temperatures as required for multi-thermal plasmas.  Including lines
provided strong constraints on dominant temperatures, but the
``abundances'' were here solely parameters, since the line flux,
continuum flux, and abundances are degenerate to a large degree.  The
result is a semi-empirical plasma model which follows the observed
continuum.

Given a continuum model, we then parametrically fit a large number of
lines (in small groups) as a sum of the plasma continuum and Gaussian
line profiles.  
%
{ 
  The free parameters were the line wavelengths and fluxes.  Since the
  lines are unresolved (the thermal broadening being below
  instrumental resolution, and with no detection of other broadening,
  such as due to turbulence, rotational, or from binary motions),
  intrinsic widths were typically frozen at a small value (e.g.,
  $2\,\mathrm{m\AA}$).  The only exception was for the Ly-$\alpha$
  lines of H-like ions which are marginally resolved doublets; and
  allowing the width to be free was sufficient for an accurate
  determination of line flux.
} 
For regions with heavily blended lines, constraints on line
separations were sometimes imposed if the features were well
identified (e.g., \eli{Ne}{10} Lyman series at $\approx
9.5$--$12.1\mang$ blending with the \eli{Mg}{11} triplet at $\approx
9.2\mang$).  Each line fit had a candidate identification with a
transition in AtomDB, based on prior experience with similar spectra
\citep[for example, see][]{Huenemoerder:Canizares:al:2003,
  Huenemoerder:01}, or if it is unknown or an unresolved blend, it may
have been fit solely to remove it from an overlapping feature of interest.
Lines with large absolute wavelength residuals relative to the AtomDB
identification were flagged as probable mis-identifications and
excluded from further analysis.  We give the line measurements for
\siggem and \hrtennn in
Tables~\ref{tbl:sgemlines}--\ref{tbl:hrtenlines}.

\begin{figure}[!htb]
  \centering\leavevmode
  \includegraphics[width=0.9\columnwidth, angle=-0, viewport=45 35 500 480]{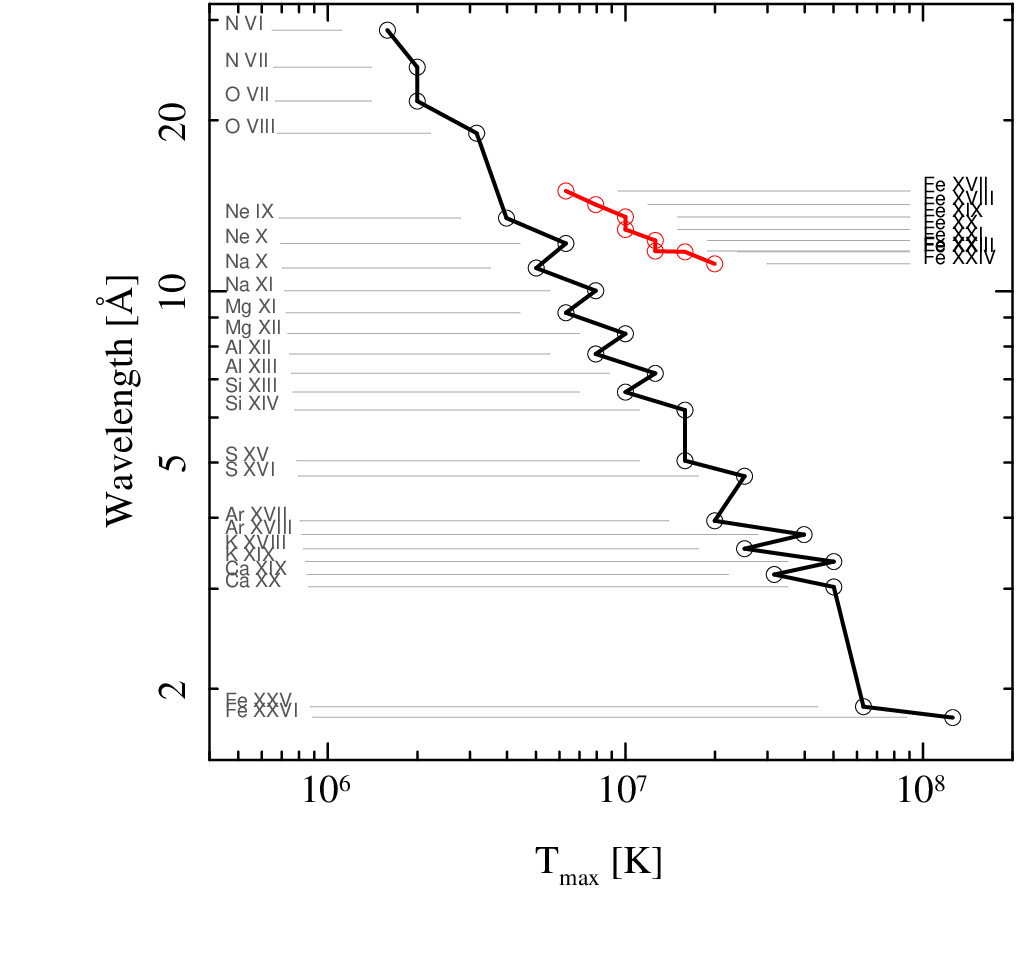}
  \caption{Over the wavelength range covered by HETG, the lines are
    sensitive to temperatures ranging from $1.5$--$60\mk$.  The circles
    mark the peak emissivity for prominent lines.  The main diagonal
    (black) shows the H- and He-like lines.  The
    \eli{Fe}{17}--\eli{Fe}{24} lines are shown as a separate curve
    (red).  }
  \label{fig:linetemps}
\end{figure}

The resulting list of lines and fluxes were then used to perform a
simultaneous determination of the EMD and elemental abundances.  This
was done by minimizing the residuals of model and measured fluxes
against a smoothness constraint on the EMD, using the emissivities
{\it vs.}\ temperature taken from AtomDB for each line.  The free
parameters were the EMD weights in each of 26 logarithmically spaced
temperature bins which spanned the sensitivity range of the lines, and
the relative abundances.  We started with a flat EMD and constant
abundances.  The constraint was cast in the form of a penalty function
proportional to the summed squared second derivative of the EMD, and
by fitting the logarithm of the EMD to enforce positivity.
Figure~\ref{fig:linetemps} shows the temperature regime covered by
lines in the \hetg bandpass ($1.5$--$25\mang$), spanning 
$1.5$--$60\mk$.  Since the continuum emissivity was not included
explicitly in this procedure, the resulting abundance and EMD
normalizations are degenerate, requiring a good starting point, and a
post facto evaluation of the binned spectrum to evaluate the continuum
level.  We performed an iteration in which we derived the EMD and
abundances, evaluated the model spectrum, scaled the overall EMD
normalization and abundances inversely, and repeated the
reconstruction until converged.
%
{ 
In \S~\ref{app:ratioresids}, we show the line flux residuals against
different parameters to demonstrate the quality of the solutions.
Ratios of observed to theoretical fluxes within a factor of 2 are
typical of such reconstructions \citep[see, for
example][]{Sanz-Forcada;Brickhouse:02}. 
} 
%

%

\begin{figure}[!htb]
  \centering\leavevmode
    \includegraphics[width=0.90\columnwidth, viewport= 50 20 515 490]{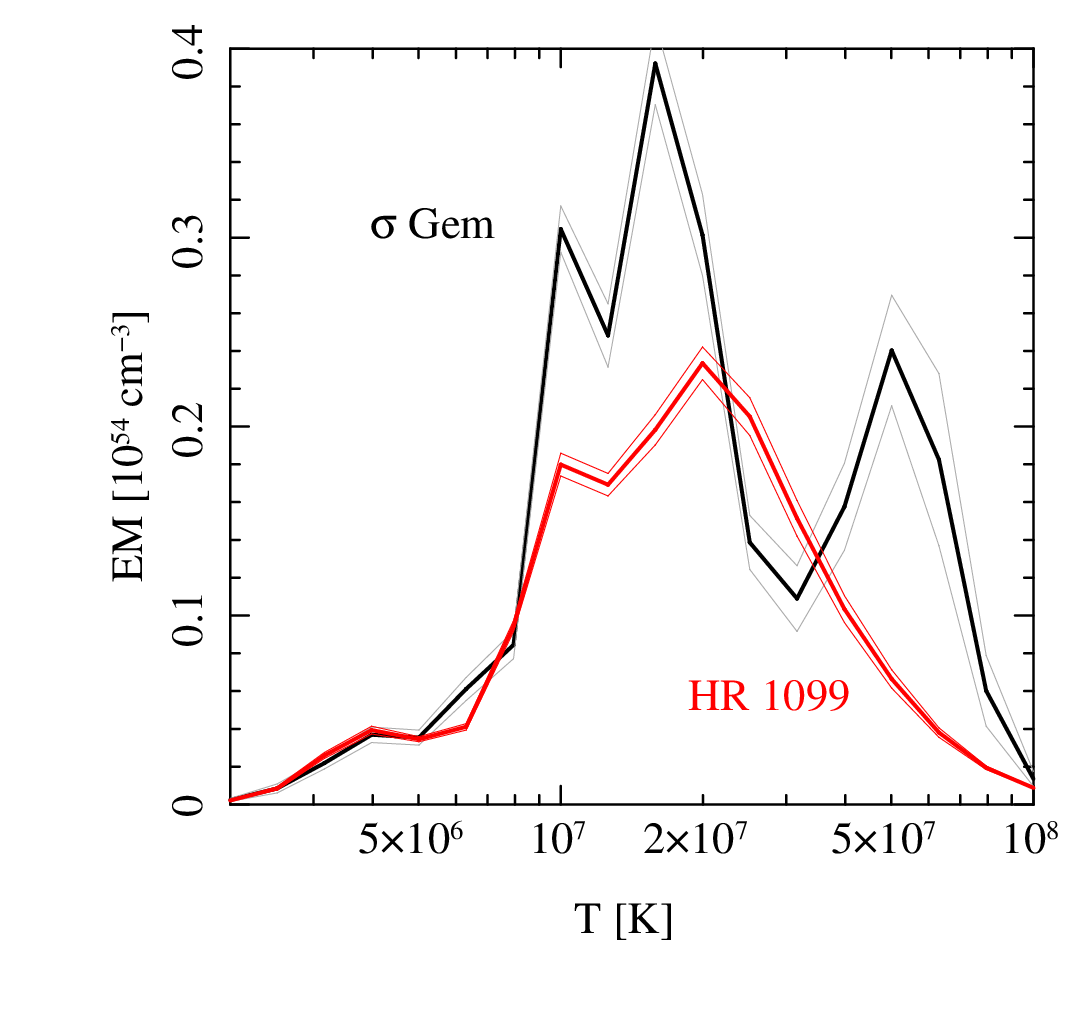}\newline
    \includegraphics[width=0.90\columnwidth, viewport= 40 20 505 480]{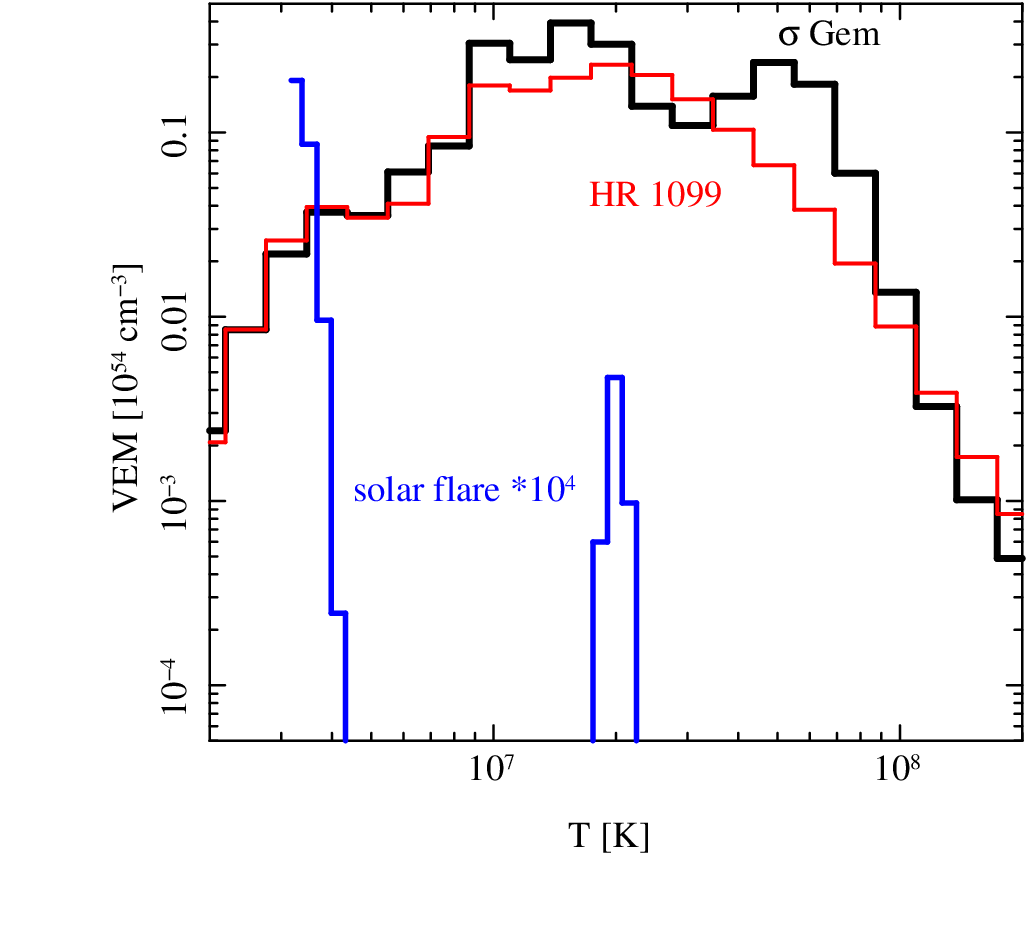}
  \caption{Top: Emission measure distributions for \siggem (black) and
    \hrtennn (red), integrated over temperature bins of 0.1 dex. The
    upper and lower boundaries are statistical     uncertainties due
    to line flux uncertainties, as determined from
    Monte-Carlo iteration.  Bottom: The same emission measure
    distributions on a logarithmic scale over ranges to allow
    inclusion of the solar flare distribution corresponding to the
    spectrum shown in Figure~\ref{fig:threespec} (blue) scaled up by a
    factor of $10^4$.
  }
  \label{fig:emd}
\end{figure}

We tested the EMD/abundance reconstruction method by fitting spectra
simulated with sums of 1--3 broken powerlaw EMDs ($EMD\propto
(T/T_0)^{\gamma(T)}$).  We used different shapes for the input EMD, from
single, narrow peaks, to 2--3 peaks of different widths, positions,
and weights.  The results showed that small wiggles in the EMD (of
order 10\%) were artifacts, that the EMD peak temperatures could be
determined to about 0.1 dex, that abundances could be determined to
about 10\% accuracy for strong-lined ions (in this self-consistent
treatment of the method), and finally that the weak line features
should have their abundances fitted post facto using the EMD solution.
The simulations also allowed us to tune the smoothness constraint,
since if it was too small, much jagged structure appeared in the
solution.  The top panel of Figure~\ref{fig:emd} gives a detailed view
of the \siggem and \hrtennn EMDs.  Values plotted are the emission
measure integrated over 0.1 dex temperature bins.  The dip in both
curves at just above $10\mk$ is probably a reconstruction artifact, as
are wiggles on the low-temperature tail.  The large peak at about
$50\mk$ in the \siggem solution, however, is required by the spectrum
(and actually required a slightly more relaxed EMD smoothness
constraint than for \hrtennn to give a good match to the model
spectrum).

%

\begin{figure}[!htb]
  \centering\leavevmode
  \includegraphics[width=0.90\columnwidth, viewport= 45 20 650 620]{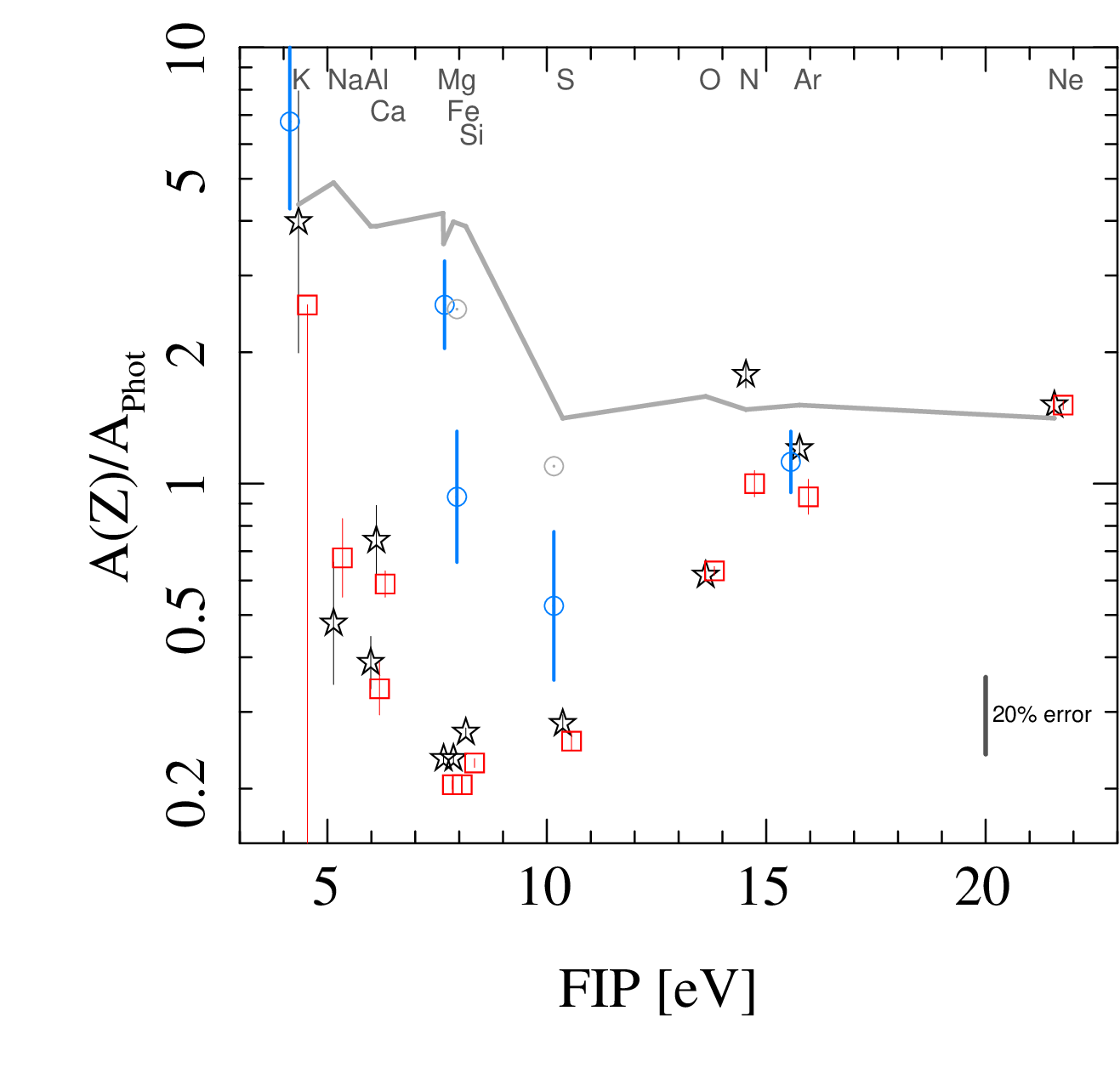}
  \caption{Abundances, relative to solar photospheric values of
    \citet{Asplund:Grevesse:al:2009} for \siggem (stars, black), and
    \hrtennn (squares, red, offset in FIP by $+0.2\,$eV).  Some error
    bars are unrealistically small, being based on emission line
    counting statistics; systematic uncertainties in atomic data and
    reconstruction methods would give about 20\% uncertainties, as
    indicated by the representative error bar in the lower right.
    Values for the particular solar flare analyzed here are shown by
    circles (blue) for K, Fe, Si, S, and Ar, offset by $-0.2\,$eV.
    The gray circles (without error bars) above the blue solar flare
    Si and S points are the isothermal-model values from prior
    analyses (see text, \S~\ref{sec:resikfits}).  The solid gray line
    is the coronal to photospheric ratio, showing the FIP effect.}
  \label{fig:abunds}
\end{figure}

After we obtained a solution, we then ran several ($\sim$100)
Monte-Carlo iterations in which we perturbed the line fluxes by their
uncertainties and then re-computed the emission measure distribution
and abundances.  This provided some idea of the uncertainty in the
solution due to counting statistics.  The $1\sigma$ envelopes are shown
in the upper panel of Figure~\ref{fig:emd}, though these do not
represent independent errors. The uncertainties from counting
statistics are relatively small.  Especially for the abundances (see
Figure~\ref{fig:abunds}), they are generally less than systematic
uncertainties expected from the underlying atomic data.

%

For lines too weak or too few to include in the EMD reconstruction, we
did post facto fits of their abundances by adopting the EMD as a frozen
quantity, then fitted only the relevant abundance in narrow regions of
the binned spectrum including the lines of interest.  We did such
for \eli{K}{18}, \eli{Na}{11}, and \eli{Al}{13}.  Abundance
values are given in Table~\ref{tbl:abund}.  We have also listed the
Monte-Carlo determined statistical uncertainties, even if
unrealistically small.  A rough estimate of the minimum uncertainty
from systematic errors is about 0.05--0.1 dex ($\sim20\%$).

%
\begin{deluxetable*}{c r cccc ccc}
  \tablecolumns{9}
  \tablewidth{0\columnwidth}
  \tablecaption{Elemental Abundances and FIP\label{tbl:abund}}
  \tabletypesize{\small}
  \tablehead{
    \colhead{Atom}&
    \colhead{FIP}&
    \colhead{AG89\tablenotemark{a}}&
    \colhead{GS98\tablenotemark{b}}&
    \colhead{Asp09\tablenotemark{c}}&
    \colhead{Solar}&
    \colhead{Solar}&
    \colhead{\siggem}&
    \colhead{\hrtennn}\\
    \colhead{}&
    \colhead{[$\ev$]}&
    \colhead{}&
    \colhead{}&
    \colhead{}&
    \colhead{Coronal\tablenotemark{d}}&
    \colhead{Flares\tablenotemark{e}}&
    \colhead{}&
    \colhead{}\\
    \colhead{(1)}&
    \colhead{(2)}&
    \colhead{(3)}&
    \colhead{(4)}&
    \colhead{(5)}&
    \colhead{(6)}&
    \colhead{(7)}&
    \colhead{(8)}&
    \colhead{(9)}
  }
  \startdata
  N&   14.534& 8.05&  7.92& 7.83& 8.00&\nodata&   8.08 (0.03)&    7.83 (0.03) \\
  O&   13.618& 8.93&  8.83& 8.69& 8.89&\nodata&   8.48 (0.01)&    8.49 (0.01) \\
  Ne&  21.564& 8.09&  8.08& 7.93& 8.08&\nodata&   8.11 (0.01)&    8.11 (0.01) \\
  Na&   5.139& 6.33&  6.33& 6.24& 6.93&\nodata&   5.92 (0.14)&    6.07 (0.09) \\
  Mg&   7.646& 7.58&  7.58& 7.60& 8.15&\nodata&   6.97 (0.01)&    6.91 (0.01) \\
  Al&   5.986& 6.47&  6.47& 6.45& 7.04&\nodata&   6.04 (0.06)&    5.98 (0.06) \\
  Si&   8.151& 7.55&  7.55& 7.51& 8.10&   7.48 (0.15) &   6.94 (0.01)&    6.87 (0.01) \\
  S&   10.360& 7.21&  7.33& 7.12& 7.27&   6.84 (0.17)&   6.57 (0.02)&    6.53 (0.02) \\
  Ar&  15.759& 6.56&  6.40& 6.40& 6.58&   6.45 (0.07)&   6.48 (0.03)&    6.37 (0.04) \\
  K&    4.341& 5.12&  5.12& 5.03& 5.67&   5.86 (0.20) &   5.63 (0.30)&    $\mathbf{<5.44}$ \\
  Ca&   6.113& 6.36&  6.36& 6.34& 6.93&\nodata&   6.21 (0.08)&    6.11 (0.03) \\
  Fe&   7.870& 7.67&  7.50& 7.50& 8.10&   7.91 (0.10)&   6.87 (0.01)&    6.81 (0.01)
  \enddata
  \tablecomments{Abundances are given on a logarithmic scale with
    $H=12$. For convenience, we list several commonly used reference
    abundances (columns 3--6).  Uncertainties on the stellar values
    (columns 8--9 in ``()'') are statistical, derived from the line
    fluxes, and do not include systematic uncertainties from atomic
    data, likely to be of order 10\%.}
  \tablenotetext{a}{\citet{Anders:89} photospheric solar values
    (default table for AtomDB \citep{Smith:01,
      Foster:Smith:Brickhouse:2012}).}
  \tablenotetext{b}{\citet{Grevesse:Sauval:1998} photospheric solar
    values (default photospheric abundances in CHIANTI
    \citep{Dere:Landi:al:2009})}
  \tablenotetext{c}{\citet{Asplund:Grevesse:al:2009}---recently
    determined, and preferred, photospheric solar abundances.}
  \tablenotetext{d}{Coronal abundances of \citet{Feldman:al:1992}, and
    of K from \citet{Landi:Feldman:Dere:2002} (a ``coronal'' table
    used in CHIANTI).}
  \tablenotetext{e}{
{
      Solar flare abundances from RESIK and
      RHESSI. Values in this column for Si, S, and Ar (with
      uncertainties) are from the re-analysis of spectra during the
      flare of 2002 December 26 (Figure~\ref{fig:threespec},
      bottom). The K abundance is from the isothermal analysis of 20
      flares \citep{Sylwester:al:2010a}, and the Fe abundance from
      analysis of 20 RHESSI flares \citep{Phillips:Dennis:2012}.
    }
  }
\end{deluxetable*}

Since EMD reconstruction results can be dependent on methods, and
since the emission measure and abundances are somewhat correlated, we
had one final diagnostic which determines abundance ratios from
linear combinations of H- and He-like ion line fluxes.  We refer to these
as temperature-insensitive (TI) abundance ratios \citep[see for
example][]{Liefke:Ness:al:2008, Garcia-Alvarez:al:2005,
  Huenemoerder:Schulz:al:2009}.  With the exception of O:Mg, we obtained
essentially the same abundance ratios from the EMD reconstruction as
from the TI-method.  We give the comparison in
Table~\ref{tbl:tinsens}.

%
\begin{deluxetable}{r rrr cc cc}
  \tablecolumns{8}
  \tablewidth{0\columnwidth}
  \tablecaption{Temperature-Insensitive Abundance Ratios\label{tbl:tinsens}}
  \tablehead{
    \colhead{Ratio}&
    \colhead{$a_0$}&
    \colhead{$a_1$}&
    \colhead{$a_2$}&
    \colhead{{\it TI}}&
    \colhead{\it EMD}&
    \colhead{\it TI}&
    \colhead{\it EMD}\\
    \colhead{(1)}&
    \colhead{(2)}&
    \colhead{(3)}&
    \colhead{(4)}&
    \colhead{(5)}&
    \colhead{(6)}&
    \colhead{(7)}&
    \colhead{(8)}
  }
  \startdata
      &\multicolumn{3}{c}{Coefficients}&\multicolumn{2}{c}{\siggem}&\multicolumn{2}{c}{\hrtennn}\\
  O:Mg&   0.340&    -6.519&     1.687&      0.7& 1.3& 0.2&  1.7\\
  Ne:O&   2.214&     1.386&    -1.723&      3.0& 3.4& 2.9&  3.0\\
  Ne:Mg&  1.270&    -0.611&     2.980&      3.8& 4.3& 4.2&  5.1\\
  Si:Mg&  0.543&     2.037&     0.0097&      1.1& 1.0& 1.1&  1.0\\
  Si:S&   0.785&     0.153&     1.570&      1.0& 1.1& 1.0&  1.0\\
  Ar:S&   2.734&     1.420&     1.897&      3.8& 3.8& 3.1&  3.0
  \enddata
  %
  \tablecomments{
{
    Abundance ratios are for abundances relative to solar
    photospheric.  E.g., Ne:Mg means
    $(A_1/A_2) = \left[A(\mathrm{Ne})/A_\odot(\mathrm{Ne})\right]
    /\left[A(\mathrm{Mg})/A_\odot(\mathrm{Mg})\right]$.
    In columns 5--8, ``{\it TI}'' means the ratio was determined from
    ``Temperature Insensitive'' line ratios, whereas ``{\it EMD}'' refers
    to values from the emission measure distribution and abundance
    reconstruction. 
    Values (and coefficients in colums 2  and 3) are referenced to
    \citet{Anders:89}.  The ratios are derived from the fluxes of the
    H-like ($F_{i,H}$) and He-like ($F_{i,He}$) resonance lines for
    elements $i=1,2$ from the coefficients via $(A_1 / A_2) = a_0 (
    F_{1,\mathrm{H}} + a_1 F_{1,\mathrm{He}} ) / ( F_{2,\mathrm{H}} +
    a_2 F_{2,\mathrm{He}} ) $.
    }
 }
\end{deluxetable}




\subsection{RESIK Fitting and Modeling}\label{sec:resikfits}

{ 
The fitting procedure for RESIK spectra, which as mentioned cover a
much smaller range (3.4--6.05~\AA) than the \hetg spectra, follows that
used in several previous analyses \citep{Sylwester:al:2010a,
  SylwesterB:al:2010, Sylwester:al:2010b, Sylwester:al:2012}. The
procedure is based on the CHIANTI database and software package
\citep{Dere:Landi:al:2009} written in Interactive Data Language (IDL),
widely used for solar x-ray and ultraviolet spectra, rather than the
one used for \hetg spectra (\S~\ref{sec:hetgspecfit}) which uses APEC
and the AtomDB atomic database. Both CHIANTI and APEC databases draw
on practically identical atomic data such as line excitation rates,
giving (as we verified) indistinguishably different $G(T_e)$ functions
for x-ray lines used in this analysis.

In previous analyses of RESIK spectra, the abundances of K, Ar, S, and
Si were estimated from solar flare and active region spectra from the
assumption that the line emission can be adequately described by an
isothermal plasma with a characteristic temperature given by the ratio
of the emission in the two wavelength channels of GOES. The lines are
mostly emitted by H-like or He-like ions but also, in the case of S
and Si, lower-temperature dielectronic lines emitted by the Li-like
ions. For all the lines analyzed, values of the line flux divided by
the volume emission measure plotted against temperature cluster about
curves having the same temperature dependence as the theoretical
$G(T_e)$ function for the line in question, as calculated from
CHIANTI. The abundance is estimated by the amount the observational
points have to be multiplied by to give the best fit to the $G(T_e)$
curve.

In recent, as yet unpublished, analysis of RESIK spectra, the
isothermal assumption has been replaced by a method in which the EMD
was derived by an iterative method relying on a Bayesian approach in
which portions of each spectrum, including lines and continuum, were
fit with a continuous function describing the EMD. This method,
described by \citet{Sylwester:Schrijver:Mewe:1980}, was used for the
analysis of non-flaring active region RESIK spectra
\citep{SylwesterB:al:2010}. The element abundances are free
parameters. While it does not follow exactly the same method as that
described for the wider-range \hetg spectra in
\S~\ref{sec:hetgspecfit}, it is equivalent in that an EMD solution
with enforced positivity and smoothness constraints are imposed with
an iterative procedure to minimize $\chi^2$ until convergence is
achieved, i.e. the EMD describes the observed spectrum to within
acceptable limits; the estimated abundances are the ones with the
least value of $\chi^2$. The RESIK spectrum shown in
Figure~\ref{fig:threespec} (bottom panel) integrated during the rise
phase of the flare on 2002 December~26 was analyzed to give the EMD
shown in Figure~\ref{fig:emd} (bottom panel); a nearly bimodal
distribution resulted in this case, with emission centered on
temperatures of $\approx 3\mk$ and $\approx 20\mk$ (other cases showed
a more continuous distribution).
This is the span of the $G(T_e)$ functions of the lines of the
principal ions (from the Li-like Si satellites to the He-like K
lines). 
The estimated abundances for Si, S, and Ar from this procedure using
several spectra during the 2002 December 26 flare are given in
Table~\ref{tbl:abund} (column 7) and in Figure~\ref{fig:abunds}.  The
Ar abundance is similar to that from the isothermal assumption
\citep{Sylwester:al:2010b}, but S and Si abundances are less. The
precise reason is under investigation, but it appears that the
assumption of a single temperature to describe RESIK specra neglects
the non-flaring active region component of the EMD which is of
importance for the relatively low-temperature S and Si lines. The K
abundance in Table~\ref{tbl:abund} is from \citet{Sylwester:al:2010a}.

The Fe abundance in Table~\ref{tbl:abund} is the average derived from
RHESSI spectra during 20 flares given by \citet{Phillips:Dennis:2012}.
} 

\subsection{Densities, Timescales, and Validity of Coronal Ionization
  Equilibrium}

Solar and stellar flares are by definition highly dynamic events which
undergo sudden heating, ionization, and recombination.  Yet we have
used plasma models in coronal ionization equilibrium (CIE), and this
requires some justification.  
The ionization and recombination times are given by
$\tau = 1/(N_e R)$ where $R$ is the rate coefficient of ionization or
recombination. Thus, $\tau$ is inversely proportional to $N_e$. 
For the \hetg stellar
spectra, we can estimate densities from the He-like triplet
forbidden-to-intercombination line ratios.
\citet{Testa:Drake:al:2004b} provide values derived from the same
\hrten data studied here, and \citet{Ness:Schmitt:al:2002} obtained
similar values for \hrten from \xmm/\rgs spectra.  Here we find
comparable values for \siggem, and we list them in
Table~\ref{tbl:ndens} along with the prior determinations for \hrten.

\begin{deluxetable}{c c c c}
  \tablecolumns{4}
  \tablewidth{0\columnwidth}
  \tablecaption{Electron Densities from He-like Triplet Ratios\label{tbl:ndens}}
  \tablehead{
    \colhead{Ion}&
    \colhead{\siggem\tablenotemark{a}}&
    \colhead{\hrten\tablenotemark{b}}&
    \colhead{\hrten\tablenotemark{c}}\\
    \colhead{(1)}&
    \colhead{(2)}&
    \colhead{(3)}&
    \colhead{(4)}
  }
  \startdata
  \eli{O}{7}&   $<10.6$&     $10.0$ $(0.6)$&      $10.4$ $(0.2)$\\
  \eli{Ne}{9}&  $<11.6$\tablenotemark{d}& \nodata& $<10.9$\\
             &  $11.0$ $(0.55)\tablenotemark{d}$&&\\
  \eli{Mg}{11}&  $<11.8$&    $12.3$ $(0.1)$&	$12.5$ $(0.5)$
  \enddata
  \tablecomments{Values are common logarithms of densities in
    $[\cmmthree]$.  The $1\sigma$ logarithmic uncertainties are given
    in parentheses.  }
  \tablenotetext{a}{Values for \siggem are from this work.}
  \tablenotetext{b}{Values from \citet{Testa:Drake:al:2004b}}
  \tablenotetext{c}{Values from \citet{Ness:Schmitt:al:2002}}
  \tablenotetext{d}{We give two values. The upper-limit used the
    emission-measure weighted line fluxes.  If we assume all emission comes
    from the temperature of maximum emissivity, then the density is
    bounded, giving the second value.}
\end{deluxetable}

For the solar flare analyzed here, we have no direct density
diagnostic, but we can be guided by previous measurements of similar
flares. \citet{Doschek:al:1981} used the \eli{O}{7} triplet in flare
spectra obtained with the {\em P78-1} spacecraft to derive values of
$N_e$ reaching $10^{12}\cmmthree$ at flare maximum, declining to about
$3 \times 10^{10}\cmmthree$ during the flare decay. Very similar
densities during flares for the higher-temperature \eli{Fe}{21} lines
(maximum of $\sim 10^{12}\cmmthree$) were derived from the Solar
Dynamics Observatory's EVE instrument by
\citet{Milligan:Kennedy:al:2012}).  At temperatures observed during
the RESIK flare, values of $R$ for both ionizations and recombinations
for ions of interest here range from
$10^{-13}\cmpthree\,\mathrm{s}\mone$ to
$10^{-11}\cmpthree\,\mathrm{s}\mone$, giving ionization and
recombination times of less than $10\,\mathrm{s}$.

Hence, we can safely assume that over the scale of the \chan stellar
observations, and the relatively gradual changes in the observed light
curves, that CIE is a reasonable assumption (a conclusion also reached
by \citet{Testa:al:2007b}, based on computations of timescales by
\citet{Golub:89}).  For solar flares similar to those seen by RESIK
and RHESSI, densities of at least $10^{11}\cmmthree$ seem to apply,
giving ionization and recombination times much less than the time
scales of the observed temperature variations.

\subsection{Prior Results}

Our results are complementary to---but in some ways discrepant
from---prior analysis of the same spectra.  \citet{Drake:01} derived
an emission measure and abundances for \hrtennn, but were concerned
primarily with relative abundances from modeling the strongest
features.  We obtain very similar abundance ratios.  Their model,
however, gives a poor representation of the observed spectrum in an
absolute sense.  Scaling their emission measure up by a factor of 8
and their abundances scaled down by a factor of 2.5 provides a
reasonable match to observed counts above $5\mang$, but is very poor
at shorter wavelengths; this is probably due to their sharp cutoff in
the EMD above $\sim25\mk$.

\citet{Nordon:Behar:2007, Nordon:Behar:2008} analyzed flares in
several stars, including \siggem and \hrtennn, using both \xmm and
\chan spectra.  The \xmm spectra do not provide high resolution data
at the short wavelengths of interest here; they relied on CCD
resolution for lines of S, Ar, and Ca.  Since the \xmm observation of
\siggem was entirely of a flare state, they used the \hetg\ \siggem
observation to represent its quiescent state.  That is rather dubious
now, given the strong, high-temperature peak in our EMD; the light
curve was only slowly descending, while a very hot EMD component is
typical of flares \citep[see for examples][]{Huenemoerder:01,
  Gudel:Audard:al:2004}.

The EMD for \siggem derived by \citet{Nordon:Behar:2008} is
qualitatively similar to ours from $5$--$30\mk$, but their high
temperature peak from $50$--$100\mk$ is many times larger.  This could
simply represent the physical reality of an extremely large and hot
flare in that observation.  Our \hrtennn EMD shape is very similar to
theirs as derived from \xmm data, despite being observed at a
different time from the \chan/\hetg\ spectra.  For the same \hetg
observation of \hrtennn given in \citet{Nordon:Behar:2007}, we have
somewhat discrepant results: while our values for $L_x$, the x-ray
luminosity over the $1$--$40\mang$ band, are of the same order
($10^{31}\lum$), our integrated emission measure ($VEM$; see
Table~\ref{tbl:stars}) is about 5 times larger.  The shape of their
EMD is also different, being broader and flatter.  Our estimate of the
$VEM$ appears robust, being obtained even if fitting a single
temperature plasma with uniformly variable abundances to a single
order of one observation---the lines match poorly overall, but the
continuum is a fair match and largely specifies the $VEM$.  
%
{ 
  The factor of 5 difference is likely due to the abundances used.
  The line luminosity is degenerate in abundance and emission measure
  which enter as a product (see equations~\ref{eqn:gt}
  and~\ref{eqn:emd}).  \citet{Nordon:Behar:2007} were interested in
  abundances changes, and did not determine the Fe abundance.  Hence,
  if one adopts an iron abundance 20\% of solar in their analysis,
  their $VEM$ would be 5 times larger.

  Another case, which at first appears discrepant, actually agrees for
  the same reason. \citet{Sanz-Forcada;Brickhouse:02} derived the EMD
  for \siggem and \hrten from UV emission lines.  They also assumed
  solar abundances, and have a $VEM$ about 5 times lower than our
  determination.  If we use our model to predict the $135.85\mang$
  Fe-blend flux, we obtain values within a factor of 2 of their
  measurements.  If we use their model to evaluate the \eli{Fe}{17}
  $15.01\mang$ flux, we get essentially the same as our measurement.
  Given that the stars are variable, we have very good agreement with
  the UV results.

  In sum, we believe our models improve on---in a global and absolute
  sense---prior works 
} 
which dealt with ratios, limited spectral ranges, or low resolution.
Our detailed modeling is especially important for determining reliable
values for some of the weak emission lines from low FIP elements,
where the multi-thermal continuum model can be very important.

\section{Discussion}

The origin of the values of elemental abundances in solar and stellar
coronae is an unsolved problem.  It is an important one because
the differences in abundances from the underlying photosphere could
provide information on the coronal structure, such as loop geometry or
Alfv\'en wave frequency and amplitude, e.g., under the ponderomotive
fractionation theory of \citet{Laming:2012, Laming:Hwang:2009}.  Or if
the relation between coronal and photospheric abundances were known,
then x-ray spectra would be very valuable in determining abundances in
star-forming regions where stars are deeply embedded and not visible
in other wavelength bands.

The x-ray spectral modeling is fairly simple; the spectra are amenable
to determination of absolute abundances, especially in the short
wavelength region where the continuum is dominated by a single
temperature, and the line-to-continuum ratio is directly proportional
to the elemental abundance.  At about $10\mk$, the continuum flux at
$3\mang$ is about 10\% that at $6\mang$, but becomes comparable by
about $30\mk$.  Our model for \siggem shows that below $6\mang$ the
primary contribution is from plasmas with $T\sim50\mk$, and $90\%$ of
the continuum is emitted by H and He thermal bremsstrahlung.  Above
$6\mang$, the continuum comes from plasmas with temperatures which
differ by more than 0.3 dex---the typical width of a line emissivity or 
$G(T)$ function---which means that a multi-thermal plasma model is
required to interpret the line-to-continuum ratio in terms of an
absolute abundance.  Hence, one must use all the lines available and
perform emission measure and abundance reconstructions.

The region below $6\mang$ fortunately contains lines from abundant
elements of both high and low FIP, in particular K ($3.35\mang$;
$FIP=4.34\ev$) and Ca ($3.0$, $3.2\mang$; $6.11\ev$), and at the other
extreme, Ar ($3.7$, $3.9\mang$; $15.76\ev$).
Hence, these abundance values are reliable and can be compared to the
solar flare values.

\begin{figure}[!htb]
  \centering\leavevmode
  \includegraphics[width=0.95\columnwidth, viewport=45 30 505 500]{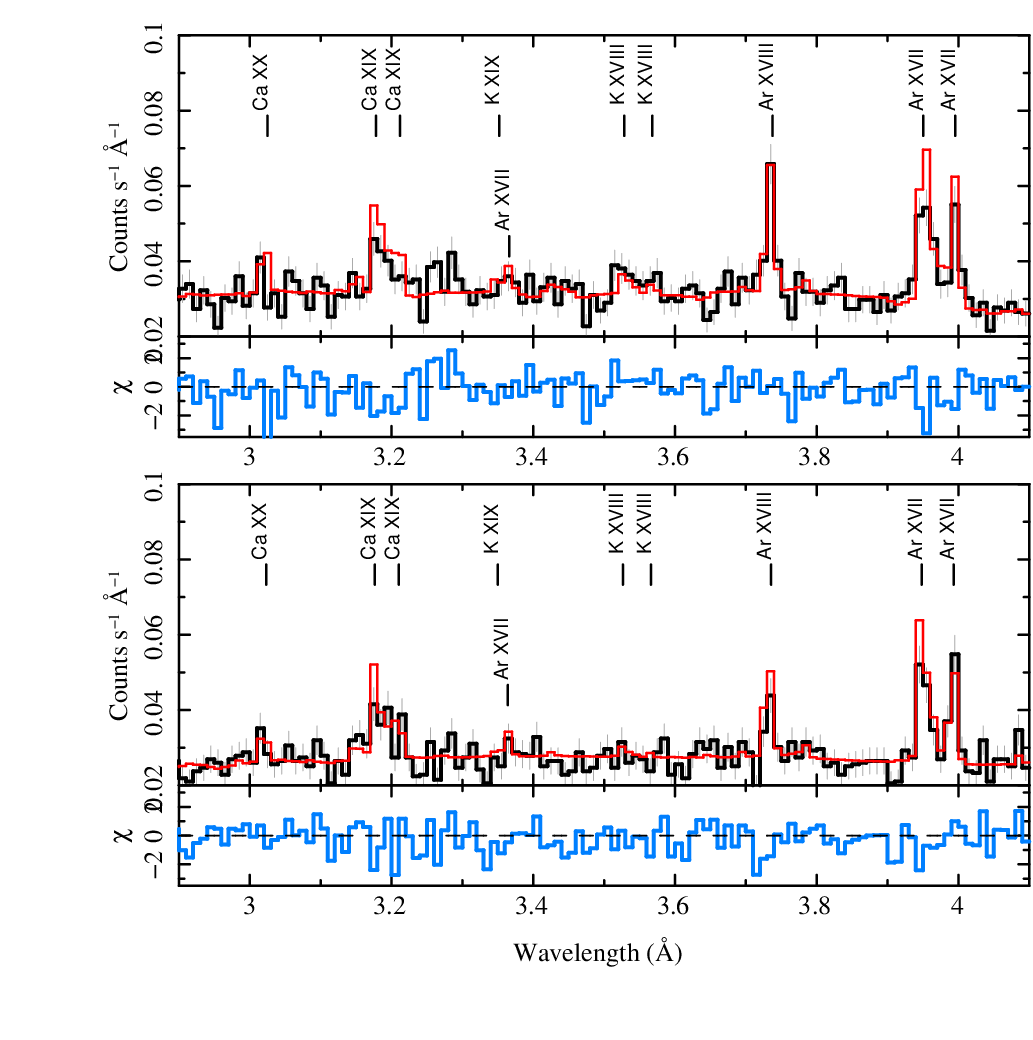}
\caption{Detail of the \eli{K}{18} region; Top: \siggem; bottom:
  \hrtennn.  Scales are the same in each panel.  Black is the observed
  spectrum, red is the convolved model, error bars are in gray, and
  below each spectrum are $\delta\chi$ residuals (in blue).}
  \label{fig:kregdetail}
\end{figure}

\begin{figure}[!htb]
  \centering\leavevmode
  \includegraphics[width=0.95\columnwidth, viewport=45 30 505 500]{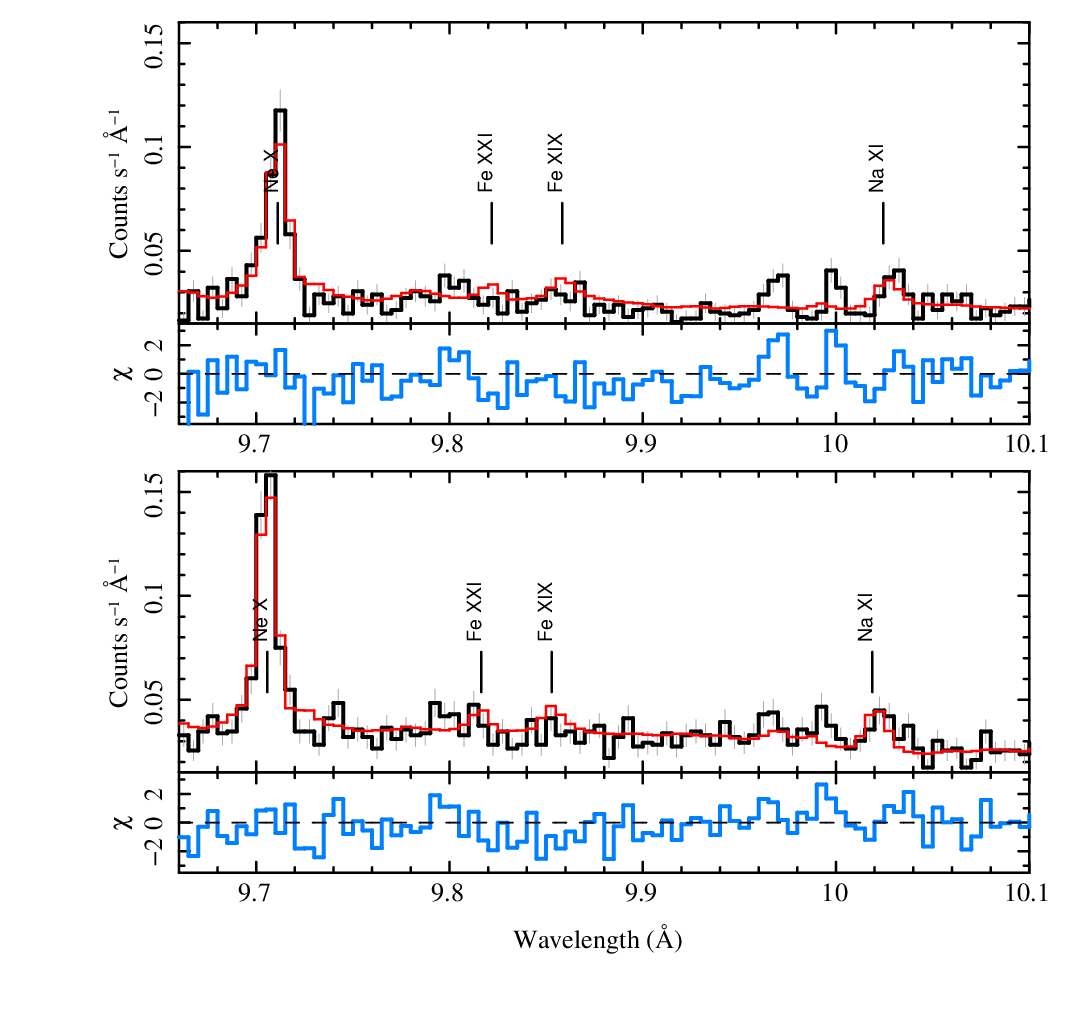}
  \caption{Detail of the \eli{Na}{11} region; Top: \siggem; bottom:
    \hrtennn.  Scales are the same in each panel.  Black is the
    observed spectrum, red is the convolved model, error bars are in gray, and
    below each spectrum are $\delta\chi$ residuals (in blue).  There are no
    identifications for the two features just blue-ward of
    \eli{Na}{11}. The strongest feature in the region is from
    \eli{Ne}{10} H-Ly$\gamma$ ($9.708\mang$).}
  \label{fig:naregdetail}
\end{figure}

Of primary interest are the species with the two lowest FIPs, K and
Na, which have not previously been measured in these stars.  The
\eli{K}{18} He-like triplet ($3.54\mang$) was marginally detected in
\siggem, but is only an upper limit in \hrtennn.  The Na abundance was
determined from the H-like \eli{Na}{11} ($10.02\mang$) line, whose
spectral region is crowded and contains some unidentified lines
(probably due to Fe), but the feature is detected at the right
wavelength for \eli{Na}{11}.  It has been seen in other coronally
active stars at about the same strength \citep{Garcia-Alvarez:al:2005,
  Sanz-Forcada:Maggio:Micela:2003}.  We show detail for these regions
in Figures~\ref{fig:kregdetail} and~\ref{fig:naregdetail}.

%

%

\begin{figure}[!htb]
  \centering\leavevmode
  \includegraphics[width=0.95\columnwidth,viewport=50 30 685 955]{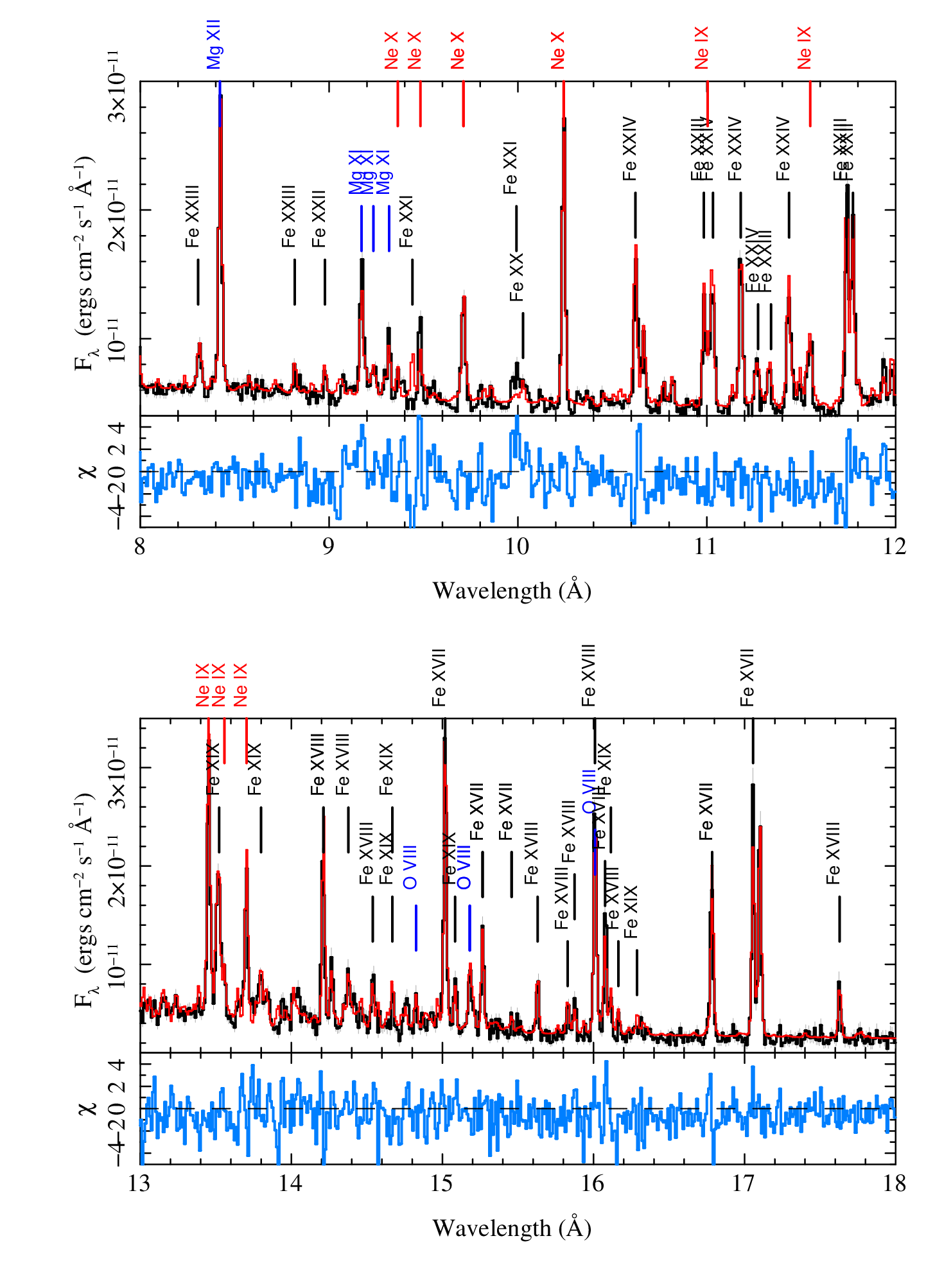}
  \caption{Here we show the \meg spectra and models for the
    $8$--$12\mang$ and $13$--$18\mang$ regions of \siggem.
    Flux-corrected spectra are in black, red is the model convolved by
    the instrumental resolution, and below each, in blue, are
    residuals.  Line identifications for the strongest emission lines
    in the regions are given.  In addition to lines from multiple ions
    of Fe, there are the \eli{Mg}{12} resonance line ($8.42\mang$) and
    \eli{Mg}{11} He-like triplet (9.17, 9.23, $9.31\mang$),
    \eli{Ne}{10} lines at 10.24, 9.78, 9.38, and $9.36\mang$, the
    \eli{Ne}{9} He-like triplet at 13.45, 13.55, and $13.70\mang$, and
    \eli{O}{8} lines at 15.176 and $16.01\mang$ (blended strongly with
    \eli{Fe}{18}).  The quality of the \hrtennn fit is similar.  }
  \label{fig:siggemspec}
\end{figure}

In the \hetg spectra, we have strong and well modeled \eli{Fe}{25}
emission ($1.85\mang$). The Fe abundance is also constrained by lower
ionization states at longer wavelengths and lower temperatures, and
appears to be robust.
%
{ 
  Emission lines from Mg, Al, Ne, O, and Ca are also present, and so
  we have derived their respective abundances.  }
In Figure~\ref{fig:siggemspec} we show a portion of the \siggem
spectrum covering a number of elements and ions to display the quality
of the fit.  The \hrtennn spectrum and model are of similar quality.

%

We show all the stellar abundances we have been able to
measure in Figure~\ref{fig:abunds}, referenced to the solar
photospheric values of \citet{Asplund:Grevesse:al:2009}.
It appears that the abundance of K in \siggem is near the solar
coronal value; that is, it is enhanced well above the solar
photospheric value.  Na, however, is much depleted, as are the other
low-FIP ($<10\ev$) elements, only attaining solar photospheric
abundances (or greater) at high FIP for N, Ar, and Ne.  In \hrtennn, K
was not detected and the upper limit is at about 2.6 times
photospheric.  Otherwise, \siggem and \hrtennn
have very similar abundances.  The important Ne:O ratio is found to be
nearly identical in \siggem and \hrtennn at 0.42, as was found for
\hrtennn by \citet{Drake:Testa:2005}.

As is typical for coronally active stars, there is no solar-like FIP
effect in which where all low-FIP elements are over-abundant relative
to the photosphere by about a factor of 4.  Instead, we see very
sub-solar abundances, except for K, N, Ar, and Ne.

The Sun also shows reduced Si and S in the 2002 December 26 flare
(blue circles in Figure~\ref{fig:abunds}); mean values obtained from
many flares (though under an isothermal assumption) are higher (gray
circles).  Abundances of Ar and K for the 2002 December 26 flare
showed no difference from the mean.  Details of the solar flare
abundance determination variance---
{we have 2795 spectra in
  various phases of 20 flares}---can be found in the series of papers
cited above (see \S~\ref{sec:resikfits}).

\begin{figure}[!htb]
  \centering\leavevmode
  \includegraphics[width=0.95\columnwidth, viewport=50 20 500 450]{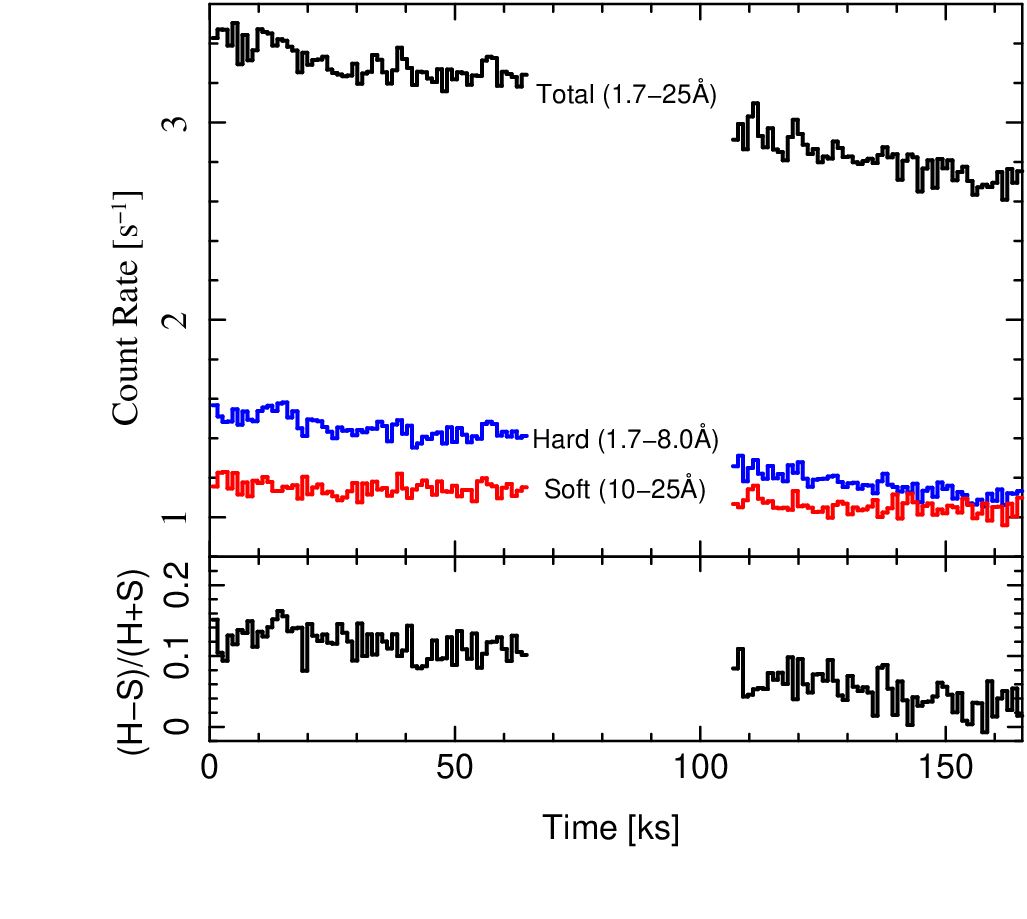}
  \caption{The light curve of \siggem derived from the \chan/\hetgs
    dispersed spectrum, summed over \heg and \meg orders $\pm1$.
    There were two separate observations made with \chan/\hetgs, each
    lasting about $60\ks$ and separated by about $50\ks$.  The total
    band (upper black) is $1.7$--$25\mang$, a hard band (blue) is
    $1.7$--$8.0\mang$, and soft (red) is $10$--$25\mang$. In the lower
    panel is a hardness ratio derived from the hard and soft band
    light curves. The decrease in hardness coincident with decreasing
    rate (proportional to emission measure) is indicative of cooling
    flares.}
  \label{fig:siggemlc}
\end{figure}

Both \siggem and \hrtennn were in flaring states during the \hetg
observations.  Light curves for the \chan\ \hrtennn data can be found
in \citet{Ayres:01} and \citet{Nordon:Behar:2007}.  In
Figure~\ref{fig:siggemlc} we show the \siggem light curve derived from
the dispersed spectral photon lists.  While there is no flare rise
seen over the times of the two \hetgs exposures, the count rate decays
steadily over the two days of observations from about $3.8\cps$ to
$3.0\cps$, and the spectrum softens as seen in a hardness ratio.  This
fading and cooling, combined with the strong, hot EMD peak, suggests
that the star is in the late stages of a large flare.

%

\section{Conclusions}

Our primary goal in this work was to compare low and high FIP
elemental abundances in the Sun and stars using x-ray spectra of
similar high-resolution and bandpass.  Such was possible using the RESIK
instrument for the Sun, and the \chan/\hetgs for stars.  In order to
obtain results for the lowest FIP element present (K), we required the
highest fluence \hetgs coronal spectra, which were of \siggem and
\hrten.  In order to have some basis for comparison of the Sun to
stars, we used a solar flare spectrum so that it has plasma at
comparable temperatures to the stars.

As a secondary goal, the richness of the \hetgs spectra allowed
determination of elemental abundances for other low FIP species,
namely Na, Al, and Ca, which had not before been measured in detail,
as well as measurements for the high-FIP elements, O, N, and Ne.

Determination of the stellar abundances from the broader \hetgs
spectrum necessarily required a full emission measure solution, since
emission measure and abundance appear as a factor in line flux
formation, and because the continuum beyond about $6\mang$ has significant
contributions from a broad range of temperatures.  These details must
be considered carefully when attempting comparisions to other work.

Hence, we have determined new elemental abundances for the lowest FIP
species---particularly K and Na, but also Ca and Al---in two stellar
coronal sources with very hot, flaring plasmas.  While the abundance
of K (having the lowest FIP) in \siggem is similar to that in the
solar corona, other low-FIP elements (Na, Al, Ca, Mg, Fe, and Si) are
strongly depleted, only becoming near or above solar for the high FIP
elements N, Ar, and Ne.  Even the stellar S abundances (considered
high-FIP) has a very low relative abundance.

Except for K, for which we only have an upper-limit in \hrtenn, the
two stellar elemental abundance distributions are remarkably similar
to each other, despite differences in their rotational periods
($2.8\,\mathrm{d}$ and $19.6\,\mathrm{d}$ for \hrten and \siggem,
respectively) and luminosity classes.

The solar flare plasma, which reaches stellar coronal temperatures of
$20\mk$, shows low-FIP elements of K and Fe to be typical of the solar
corona (that is, enhanced relative to photospheric values), while Si
and S are photospheric or lower, but still above the stellar values.
The abundances of Ar is similar in the solar and stellar flares.  The
overall trend with FIP is similar to, but less extreme than the
stellar case, with the exception of Fe having high relative abundance,
though it has about the same FIP as Si.

Both the stellar and solar cases show abundance trends more
complicated than simple FIP or ``inverse''-FIP during flares.  Whether
other active stars have similar patterns at the lowest FIP would
require investment of $100$-$200\ks$ per star to raise some of the
existing lower fluence \hetgs observations of coronal sources to
comparable levels.  Likewise, further solar flare analysis is required
to determine whether the Fe and Si abundances are always so different,
or whether the solar trend sometimes mimics the stellar trend more
closely.




\acknowledgements Acknowledgements: Support for this work was provided
by the National Aeronautics and Space Administration through the
Smithsonian Astrophysical Observatory contract SV3-73016 to MIT for
Support of the Chandra X-Ray Center, which is operated by the
Smithsonian Astrophysical Observatory for and on behalf of the
National Aeronautics Space Administration under contract NAS8-03060
(DPH).  We also gratefully acknowledge financial support from the
Polish Ministry of Education and Science (Grant 2011/01/B/ST9/05861),
the European Commissions Seventh Framework Programme (FP7/2007-2013)
under the grant agreement eHEROES (project no. 284461), and the UK
Royal Society/Polish Academy of Sciences International Joint Project
for travel support. The RESIK spectra were analyzed with CHIANTI which
is a collaborative project involving George Mason University, the
University of Michigan, and the University of Cambridge.

{\it Facilities:} \facility{ CXO (HETG/ACIS)}, \facility{{\it{CORONAS-F}} (RESIK)}

%

\begin{thebibliography}{}

\bibitem[\protect\astroncite{{Anders} \& {Grevesse}}{1989}]{Anders:89}
{Anders}, E., \& {Grevesse}, N.,  1989, \gca, 53, 197

\bibitem[\protect\astroncite{{Asplund} et~al.}{2009}]{Asplund:Grevesse:al:2009}
{Asplund}, M., {Grevesse}, N., {Sauval}, A.~J., \& {Scott}, P.,  2009, \araa,
  47, 481

\bibitem[\protect\astroncite{{Audard}, {G{\" u}del} \&
  {Mewe}}{2001}]{Audard:01a}
{Audard}, M., {G{\" u}del}, M., \& {Mewe}, R.,  2001, \aap, 365, L318

\bibitem[\protect\astroncite{{Ayres} et~al.}{2001}]{Ayres:01}
{Ayres}, T.~R., {Brown}, A., {Osten}, R.~A., {Huenemoerder}, D.~P., {Drake},
  J.~J., {Brickhouse}, N.~S., \& {Linsky}, J.~L.,  2001, \apj, 549, 554

\bibitem[\protect\astroncite{{Brinkman} et~al.}{2001}]{Brinkman:01}
{Brinkman}, A.~C., et~al., 2001, \aap, 365, L324

\bibitem[\protect\astroncite{{Canizares} et~al.}{2005}]{HETG:2005}
{Canizares}, C.~R., et~al., 2005, \pasp, 117, 1144

\bibitem[\protect\astroncite{{Davis}}{2001}]{DavisJE:2001b}
{Davis}, J.~E.,  2001, \apj, 548, 1010

\bibitem[\protect\astroncite{{Dere} et~al.}{2009}]{Dere:Landi:al:2009}
{Dere}, K.~P., {Landi}, E., {Young}, P.~R., {Del Zanna}, G., {Landini}, M., \&
  {Mason}, H.~E.,  2009, \aap, 498, 915

\bibitem[\protect\astroncite{{Doschek} et~al.}{1981}]{Doschek:al:1981}
{Doschek}, G.~A., {Feldman}, U., {Landecker}, P.~B., \& {McKenzie}, D.~L.,
  1981, \apj, 249, 372

\bibitem[\protect\astroncite{{Drake} et~al.}{2001}]{Drake:01}
{Drake}, J.~J., {Brickhouse}, N.~S., {Kashyap}, V., {Laming}, J.~M.,
  {Huenemoerder}, D.~P., {Smith}, R., \& {Wargelin}, B.~J.,  2001, \apjl, 548,
  L81

\bibitem[\protect\astroncite{{Drake} \& {Testa}}{2005}]{Drake:Testa:2005}
{Drake}, J.~J., \& {Testa}, P.,  2005, \nat, 436, 525

\bibitem[\protect\astroncite{{Eker} et~al.}{2008}]{CABS3}
{Eker}, Z., et~al., 2008, \mnras, 389, 1722

\bibitem[\protect\astroncite{{Feldman}}{1992}]{Feldman:92}
{Feldman}, U.,  1992, \physscr, 46, 202

\bibitem[\protect\astroncite{{Feldman} \& {Laming}}{2000}]{Feldman:Laming:00}
{Feldman}, U., \& {Laming}, J.,  2000, \physscr, 61, 222

\bibitem[\protect\astroncite{{Feldman} et~al.}{1992}]{Feldman:al:1992}
{Feldman}, U., {Mandelbaum}, P., {Seely}, J.~F., {Doschek}, G.~A., \& {Gursky},
  H.,  1992, \apjs, 81, 387

\bibitem[\protect\astroncite{{Foster}
  et~al.}{2012}]{Foster:Smith:Brickhouse:2012}
{Foster}, A.~R., {Ji}, L., {Smith}, R.~K., \& {Brickhouse}, N.~S.,  2012, \apj,
  756, 128

\bibitem[\protect\astroncite{{Fruscione} et~al.}{2006}]{CIAO:2006}
{Fruscione}, A., et~al., 2006,
\newblock in SPIE Conference Series, Vol. 6270

\bibitem[\protect\astroncite{{G{\" u}del} et~al.}{2004}]{Gudel:Audard:al:2004}
{G{\" u}del}, M., {Audard}, M., {Reale}, F., {Skinner}, S.~L., \& {Linsky},
  J.~L.,  2004, \aap, 416, 713

\bibitem[\protect\astroncite{{Garc{\'{\i}}a-Alvarez}
  et~al.}{2005}]{Garcia-Alvarez:al:2005}
{Garc{\'{\i}}a-Alvarez}, D., {Drake}, J.~J., {Lin}, L., {Kashyap}, V.~L., \&
  {Ball}, B.,  2005, \apj, 621, 1009

\bibitem[\protect\astroncite{{Golub}, {Hartquist} \&
  {Quillen}}{1989}]{Golub:89}
{Golub}, L., {Hartquist}, T.~W., \& {Quillen}, A.~C.,  1989, \solphys, 122, 245

\bibitem[\protect\astroncite{{Grevesse} \&
  {Sauval}}{1998}]{Grevesse:Sauval:1998}
{Grevesse}, N., \& {Sauval}, A.~J.,  1998, Space Science Reviews, 85, 161

\bibitem[\protect\astroncite{{Huenemoerder}
  et~al.}{2003}]{Huenemoerder:Canizares:al:2003}
{Huenemoerder}, D.~P., {Canizares}, C.~R., {Drake}, J.~J., \& {Sanz-Forcada},
  J.,  2003, \apj, 595, 1131

\bibitem[\protect\astroncite{{Huenemoerder}, {Canizares} \&
  {Schulz}}{2001}]{Huenemoerder:01}
{Huenemoerder}, D.~P., {Canizares}, C.~R., \& {Schulz}, N.~S.,  2001, \apj,
  559, 1135

\bibitem[\protect\astroncite{{Huenemoerder}
  et~al.}{2009}]{Huenemoerder:Schulz:al:2009}
{Huenemoerder}, D.~P., {Schulz}, N.~S., {Testa}, P., {Kesich}, A., \&
  {Canizares}, C.~R.,  2009, \apj, 707, 942

\bibitem[\protect\astroncite{{Jordan}}{1975}]{Jordan:1975}
{Jordan}, C.,  1975, \mnras, 170, 429

\bibitem[\protect\astroncite{{Kaastra} et~al.}{1996}]{Kaastra:96}
{Kaastra}, J.~S., {Mewe}, R., {Liedahl}, D.~A., {Singh}, K.~P., {White}, N.~E.,
  \& {Drake}, S.~A.,  1996, \aap, 314, 547

\bibitem[\protect\astroncite{{Laming}}{2012}]{Laming:2012}
{Laming}, J.~M.,  2012, \apj, 744, 115

\bibitem[\protect\astroncite{{Laming} \& {Hwang}}{2009}]{Laming:Hwang:2009}
{Laming}, J.~M., \& {Hwang}, U.,  2009, \apjl, 707, L60

\bibitem[\protect\astroncite{{Landi}, {Feldman} \&
  {Dere}}{2002}]{Landi:Feldman:Dere:2002}
{Landi}, E., {Feldman}, U., \& {Dere}, K.~P.,  2002, \apjs, 139, 281

\bibitem[\protect\astroncite{{Lemen} et~al.}{1989}]{Lemen:1989}
{Lemen}, J.~R., {Mewe}, R., {Schrijver}, C.~J., \& {Fludra}, A.,  1989, \apj,
  341, 474

\bibitem[\protect\astroncite{{Liefke}, {Fuhrmeister} \&
  {Schmitt}}{2010}]{Liefke:al:2010}
{Liefke}, C., {Fuhrmeister}, B., \& {Schmitt}, J.~H.~M.~M.,  2010, \aap, 514,
  A94

\bibitem[\protect\astroncite{{Liefke} et~al.}{2008}]{Liefke:Ness:al:2008}
{Liefke}, C., {Ness}, J.~., {Schmitt}, J.~H.~M.~M., \& {Maggio}, A.,  2008,
  \aap, 491, 859

\bibitem[\protect\astroncite{{Mazzotta} et~al.}{1998}]{Mazzotta:98}
{Mazzotta}, P., {Mazzitelli}, G., {Colafrancesco}, S., \& {Vittorio}, N.,
  1998, \aaps, 133, 403

\bibitem[\protect\astroncite{{Milligan}
  et~al.}{2012}]{Milligan:Kennedy:al:2012}
{Milligan}, R.~O., {Kennedy}, M.~B., {Mathioudakis}, M., \& {Keenan}, F.~P.,
  2012, \apjl, 755, L16

\bibitem[\protect\astroncite{{Ness} et~al.}{2002}]{Ness:Schmitt:al:2002}
{Ness}, J.-U., {Schmitt}, J.~H.~M.~M., {Burwitz}, V., {Mewe}, R., {Raassen},
  A.~J.~J., {van der Meer}, R.~L.~J., {Predehl}, P., \& {Brinkman}, A.~C.,
  2002, \aap, 394, 911

\bibitem[\protect\astroncite{{Nordon} \& {Behar}}{2007}]{Nordon:Behar:2007}
{Nordon}, R., \& {Behar}, E.,  2007, \aap, 464, 309

\bibitem[\protect\astroncite{{Nordon} \& {Behar}}{2008}]{Nordon:Behar:2008}
{Nordon}, R., \& {Behar}, E.,  2008, \aap, 482, 639

\bibitem[\protect\astroncite{{Phillips} \&
  {Dennis}}{2012}]{Phillips:Dennis:2012}
{Phillips}, K.~J.~H., \& {Dennis}, B.~R.,  2012, \apj, 748, 52

\bibitem[\protect\astroncite{{Phillips}
  et~al.}{2010}]{Phillips:Sylwester:al:2010}
{Phillips}, K.~J.~H., {Sylwester}, J., {Sylwester}, B., \& {Kuznetsov}, V.~D.,
  2010, \apj, 711, 179

\bibitem[\protect\astroncite{{Pottasch}}{1963}]{Pottasch:1963}
{Pottasch}, S.~R.,  1963, \apj, 137, 945

\bibitem[\protect\astroncite{{Sanz-Forcada}, {Brickhouse} \&
  {Dupree}}{2002}]{Sanz-Forcada;Brickhouse:02}
{Sanz-Forcada}, J., {Brickhouse}, N.~S., \& {Dupree}, A.~K.,  2002, \apj, 570,
  799

\bibitem[\protect\astroncite{{Sanz-Forcada}, {Maggio} \&
  {Micela}}{2003}]{Sanz-Forcada:Maggio:Micela:2003}
{Sanz-Forcada}, J., {Maggio}, A., \& {Micela}, G.,  2003, \aap, 408, 1087

\bibitem[\protect\astroncite{{Smith} et~al.}{2001}]{Smith:01}
{Smith}, R.~K., {Brickhouse}, N.~S., {Liedahl}, D.~A., \& {Raymond}, J.~C.,
  2001, \apjl, 556, L91

\bibitem[\protect\astroncite{{Strassmeier}}{2009}]{Strassmeier:2009}
{Strassmeier}, K.~G.,  2009, \aapr, 17, 251

\bibitem[\protect\astroncite{{Sylwester}, {Sylwester} \&
  {Phillips}}{2010}]{SylwesterB:al:2010}
{Sylwester}, B., {Sylwester}, J., \& {Phillips}, K.~J.~H.,  2010, \aap, 514,
  A82

\bibitem[\protect\astroncite{{Sylwester} et~al.}{2005}]{Sylwester:al:2005}
{Sylwester}, J., et~al., 2005, \solphys, 226, 45

\bibitem[\protect\astroncite{{Sylwester}, {Schrijver} \&
  {Mewe}}{1980}]{Sylwester:Schrijver:Mewe:1980}
{Sylwester}, J., {Schrijver}, J., \& {Mewe}, R.,  1980, \solphys, 67, 285

\bibitem[\protect\astroncite{{Sylwester} et~al.}{2010a}]{Sylwester:al:2010b}
{Sylwester}, J., {Sylwester}, B., {Phillips}, K.~J.~H., \& {Kuznetsov}, V.~D.,
  2010a, \apj, 720, 1721

\bibitem[\protect\astroncite{{Sylwester} et~al.}{2010b}]{Sylwester:al:2010a}
{Sylwester}, J., {Sylwester}, B., {Phillips}, K.~J.~H., \& {Kuznetsov}, V.~D.,
  2010b, \apj, 710, 804

\bibitem[\protect\astroncite{{Sylwester} et~al.}{2012}]{Sylwester:al:2012}
{Sylwester}, J., {Sylwester}, B., {Phillips}, K.~J.~H., \& {Kuznetsov}, V.~D.,
  2012, \apj, 751, 103

\bibitem[\protect\astroncite{{Testa}, {Drake} \&
  {Peres}}{2004}]{Testa:Drake:al:2004b}
{Testa}, P., {Drake}, J.~J., \& {Peres}, G.,  2004, \apj, 617, 508

\bibitem[\protect\astroncite{{Testa} et~al.}{2007}]{Testa:al:2007b}
{Testa}, P., {Drake}, J.~J., {Peres}, G., \& {Huenemoerder}, D.~P.,  2007,
  \apj, 665, 1349

\bibitem[\protect\astroncite{{van Leeuwen}}{2007}]{vanLeeuwen:2007}
{van Leeuwen}, F.,  2007, \aap, 474, 653

\bibitem[\protect\astroncite{{Wood}, {Laming} \&
  {Karovska}}{2012}]{Wood:al:2012}
{Wood}, B.~E., {Laming}, J.~M., \& {Karovska}, M.,  2012, \apj, 753, 76

\bibitem[\protect\astroncite{{Wood} \& {Linsky}}{2010}]{Wood:Linsky:2010}
{Wood}, B.~E., \& {Linsky}, J.~L.,  2010, \apj, 717, 1279

\end{thebibliography}

\clearpage

\LongTables

\begin{deluxetable}{r c r r r r}
  \tablecolumns{6}
  \tablewidth{0.8\columnwidth}
  \tablecaption{\siggem Line Measurements\label{tbl:sgemlines}}
  %
  \tablehead{
    \colhead{Ion}&
    \colhead{$\log T_\mathrm{max}$\tablenotemark{a}}&
    \colhead{$\lambda_0$\tablenotemark{b}}&
    \colhead{$\lambda_\mathrm{obs}$\tablenotemark{c}}&
    \colhead{$f$\tablenotemark{d}}&
    \colhead{$f_\mathrm{model}$\tablenotemark{e}}\\
    \colhead{}&
    \colhead{$\log\,$[K]}&
    \colhead{[$\rmang$]}&
    \colhead{[$\rmang\,$($\mathrm{m}\mang$)]}&
    \multicolumn{2}{c}{[$10^{-6}\apflux$]}\\
    \colhead{(1)}&
    \colhead{(2)}&
    \colhead{(3)}&
    \colhead{(4)}&
    \colhead{(5)}&
    \colhead{(6)}
  }
  \startdata
\eli{Fe}{25}&  7.82&   1.8607&    1.858 (1.4)&     83.4 (8.0)&     68.5\\ 
\eli{Ca}{19}&  7.50&   3.1772&    3.176 (2.0)&     16.8 (3.2)&     13.5\\ 
\eli{Ca}{19}&  7.46&   3.1909&    3.192 (3.6)&      9.8 (2.9)&      4.4\\ 
\eli{Ca}{19}&  7.46&   3.2110&    3.210 (3.4)&      7.4 (2.5)&      4.5\\ 
\eli{K}{18}&  7.42&   3.5273&    3.521 (5.2)&      6.8 (2.5)&      0.8\\ 
\eli{K}{18}&  7.36&   3.5434&    3.545 (0.0)&      2.3 (2.3)&      0.1\\ 
\eli{K}{18}&  7.39&   3.5669&    3.568 (0.0)&      6.0 (2.5)&      0.4\\ 
\eli{Ar}{18}&  7.73&   3.7338&    3.736 (0.7)&     28.4 (2.9)&     29.5\\ 
\eli{Ar}{17}&  7.36&   3.9491&    3.950 (0.6)&     38.7 (3.4)&     36.8\\ 
\eli{Ar}{17}&  7.31&   3.9676&    3.968 (1.1)&     17.8 (2.9)&     10.0\\ 
\eli{S}{16}&  7.51&   3.9923&    3.995 (0.9)&     25.1 (3.0)&     11.8\\ 
\eli{S}{16}&  7.57&   4.7301&    4.731 (0.6)&     47.3 (3.8)&     52.4\\ 
\eli{Si}{14}&  7.43&   4.9462&    4.948 (1.9)&      8.4 (2.9)&      4.3\\ 
\eli{S}{15}&  7.20&   5.0387&    5.041 (0.5)&     58.4 (4.2)&     56.3\\ 
\eli{S}{15}&  7.16&   5.0648&    5.068 (2.0)&     13.2 (3.2)&     12.8\\ 
\eli{S}{15}&  7.17&   5.1015&    5.101 (1.0)&     31.3 (3.7)&     24.7\\ 
\eli{Si}{14}&  7.42&   5.2174&    5.220 (0.9)&     26.2 (3.8)&     24.1\\ 
\eli{Si}{13}&  7.07&   5.6805&    5.685 (1.3)&     17.1 (3.1)&     16.9\\ 
\eli{Si}{14}&  7.40&   6.1831&    6.185 (0.2)&    162.5 (3.9)&    164.8\\ 
\eli{Si}{13}&  7.03&   6.6479&    6.651 (0.2)&    125.5 (3.5)&    120.2\\ 
\eli{Si}{13}&  6.99&   6.6866&    6.690 (0.6)&     26.8 (2.4)&     22.9\\ 
\eli{Si}{13}&  7.00&   6.7403&    6.742 (0.3)&     82.0 (2.9)&     58.2\\ 
\eli{Mg}{12}&  7.22&   7.1063&    7.109 (0.6)&     26.7 (2.3)&     30.6\\ 
\eli{Al}{13}&  7.38&   7.1714&    7.172 (0.7)&     30.0 (2.4)&     29.0\\ 
\eli{Al}{12}&  6.94&   7.7573&    7.761 (1.1)&     16.1 (2.0)&      8.8\\ 
\eli{Mg}{11}&  6.87&   7.8503&    7.853 (1.3)&     12.3 (2.1)&     11.5\\ 
\eli{Fe}{24}&  7.46&   7.9857&    7.986 (0.9)&     22.2 (2.4)&     19.0\\ 
\eli{Fe}{24}&  7.46&   7.9960&    7.998 (1.1)&     12.3 (2.4)&      9.5\\ 
\eli{Fe}{23}&  7.28&   8.3038&    8.307 (1.3)&     17.8 (3.2)&     19.4\\ 
\eli{Fe}{24}&  7.44&   8.3161&    8.320 (0.9)&     26.4 (3.4)&     19.1\\ 
\eli{Fe}{24}&  7.44&   8.3761&    8.378 (2.0)&      9.0 (2.5)&      7.7\\ 
\eli{Mg}{12}&  7.19&   8.4219&    8.424 (0.2)&    202.8 (5.3)&    208.2\\ 
\eli{Fe}{21}&  7.10&   8.5740&    8.578 (1.2)&     13.3 (2.3)&     11.2\\ 
\eli{Fe}{23}&  7.27&   8.8149&    8.819 (0.8)&     24.0 (2.5)&     22.0\\ 
\eli{Fe}{22}&  7.17&   8.9748&    8.979 (0.8)&     21.2 (2.4)&     20.8\\ 
\eli{Mg}{11}&  6.84&   9.1687&    9.172 (0.3)&    112.1 (4.1)&     79.0\\ 
\eli{Fe}{21}&  7.10&   9.1944&    9.192 (1.9)&     20.4 (2.9)&      8.4\\ 
\eli{Mg}{11}&  6.80&   9.2297&    9.233 (1.3)&     19.7 (2.8)&     12.7\\ 
\eli{Mg}{11}&  6.81&   9.3143&    9.318 (0.4)&     56.5 (3.1)&     39.5\\ 
\eli{Ne}{10}&  6.99&   9.4808&    9.481 (0.4)&     64.0 (2.7)&     38.9\\ 
\eli{Fe}{19}&  6.97&   9.6951&    9.698 (2.7)&     18.1 (3.8)&      7.6\\ 
\eli{Ne}{10}&  6.98&   9.7082&    9.712 (0.7)&     90.5 (4.8)&     89.7\\ 
\eli{Na}{11}&  7.08&  10.0240&   10.031 (1.0)&     26.9 (3.1)&     23.2\\ 
\eli{Ni}{19}&  6.87&  10.1100&   10.112 (4.0)&      9.7 (4.7)&      3.9\\ 
\eli{Ne}{10}&  6.97&  10.2390&   10.242 (0.0)&    260.3 (6.2)&    252.8\\ 
\eli{Fe}{24}&  7.45&  10.6190&   10.625 (0.5)&    145.8 (5.4)&    134.3\\ 
\eli{Fe}{24}&  7.45&  10.6630&   10.665 (0.5)&     69.9 (4.2)&     69.4\\ 
\eli{Fe}{19}&  6.97&  10.8160&   10.823 (1.0)&     29.3 (3.3)&     27.1\\ 
\eli{Fe}{23}&  7.27&  10.9810&   10.985 (0.0)&     93.1 (4.8)&    104.8\\ 
\eli{Ne}{9}&  6.66&  11.0010&   11.005 (1.5)&     35.6 (4.2)&     22.1\\ 
\eli{Fe}{23}&  7.27&  11.0190&   11.024 (1.0)&     76.3 (7.2)&     67.2\\ 
\eli{Fe}{24}&  7.42&  11.0290&   11.036 (0.5)&     72.9 (7.3)&     88.3\\ 
\eli{Fe}{17}&  6.76&  11.1310&   11.138 (2.0)&     18.9 (3.8)&     19.8\\ 
\eli{Fe}{24}&  7.42&  11.1760&   11.179 (0.0)&    166.7 (6.7)&    159.2\\ 
\eli{Fe}{17}&  6.76&  11.2540&   11.255 (1.5)&     24.6 (4.9)&     27.4\\ 
\eli{Fe}{24}&  7.42&  11.2680&   11.268 (1.0)&     46.9 (5.2)&     36.5\\ 
\eli{Fe}{18}&  6.89&  11.3260&   11.330 (0.5)&     48.1 (4.0)&     38.7\\ 
\eli{Fe}{18}&  6.88&  11.5270&   11.531 (1.0)&     48.9 (4.4)&     37.8\\ 
\eli{Ne}{9}&  6.64&  11.5440&   11.551 (0.5)&     68.7 (4.8)&     63.6\\ 
\eli{Fe}{23}&  7.26&  11.7360&   11.744 (0.5)&    238.8 (7.4)&    223.9\\ 
\eli{Fe}{22}&  7.16&  11.7700&   11.775 (0.0)&    209.9 (7.0)&    188.6\\ 
\eli{Fe}{22}&  7.15&  11.9320&   11.937 (1.0)&     65.7 (4.9)&     27.2\\ 
\eli{Ne}{10}&  6.94&  12.1350&   12.137 (3.5)&   1839.0 (19.3)&   1838.2\\ 
\eli{Fe}{23}&  7.25&  12.1610&   12.162 (1.0)&    120.4 (9.2)&    119.6\\ 
\eli{Fe}{17}&  6.75&  12.2660&   12.268 (1.0)&     70.9 (9.7)&     60.0\\ 
\eli{Fe}{21}&  7.09&  12.2840&   12.289 (0.5)&    304.5 (12.2)&    342.4\\ 
\eli{Fe}{22}&  7.15&  12.7540&   12.756 (0.5)&     79.9 (6.6)&     64.5\\ 
\eli{Fe}{20}&  7.03&  12.8240&   12.831 (1.0)&    140.1 (10.0)&     60.1\\ 
\eli{Fe}{20}&  7.03&  12.8460&   12.849 (1.0)&    123.9 (14.1)&    140.4\\ 
\eli{Fe}{20}&  7.03&  13.3850&   13.381 (2.0)&     47.1 (9.2)&     33.0\\ 
\eli{Fe}{19}&  6.96&  13.4230&   13.434 (2.5)&     42.2 (7.9)&     24.1\\ 
\eli{Ne}{9}&  6.61&  13.4470&   13.452 (0.5)&    414.6 (14.3)&    443.3\\ 
\eli{Fe}{19}&  6.96&  13.4620&   13.471 (1.0)&     92.5 (8.7)&     55.8\\ 
\eli{Fe}{19}&  6.96&  13.5180&   13.527 (0.5)&    188.4 (12.1)&    211.4\\ 
\eli{Ne}{9}&  6.58&  13.5520&   13.558 (1.0)&    101.5 (9.2)&     71.7\\ 
\eli{Fe}{19}&  6.96&  13.6450&   13.654 (1.5)&     55.9 (6.9)&     33.8\\ 
\eli{Ne}{9}&  6.59&  13.6990&   13.703 (0.5)&    279.7 (11.5)&    249.3\\ 
\eli{Fe}{20}&  7.02&  13.7670&   13.771 (2.5)&     54.8 (7.3)&     23.2\\ 
\eli{Fe}{19}&  6.96&  13.7950&   13.800 (1.0)&    102.6 (8.6)&     90.1\\ 
\eli{Fe}{18}&  6.87&  14.2080&   14.209 (0.5)&    309.1 (14.3)&    387.5\\ 
\eli{Fe}{18}&  6.87&  14.2560&   14.263 (1.0)&     87.1 (7.1)&     76.7\\ 
\eli{Fe}{20}&  7.02&  14.2670&   14.276 (1.5)&     54.0 (12.1)&     39.3\\ 
\eli{Fe}{18}&  6.87&  14.3430&   14.350 (1.5)&     52.1 (7.9)&     43.0\\ 
\eli{Fe}{18}&  6.87&  14.3730&   14.379 (1.0)&    112.8 (10.0)&     92.5\\ 
\eli{Fe}{18}&  6.87&  14.5340&   14.541 (1.0)&    115.8 (12.8)&     76.8\\ 
\eli{Fe}{19}&  6.96&  14.6640&   14.671 (1.0)&     72.4 (9.6)&     85.0\\ 
\eli{O}{8}&  6.71&  14.8210&   14.823 (1.0)&     60.4 (8.5)&     46.9\\ 
\eli{Fe}{17}&  6.73&  15.0140&   15.017 (0.5)&    542.6 (18.0)&    563.4\\ 
\eli{Fe}{19}&  6.95&  15.0790&   15.085 (1.0)&    118.9 (9.5)&     87.4\\ 
\eli{O}{8}&  6.70&  15.1760&   15.180 (1.0)&    118.1 (9.3)&    106.7\\ 
\eli{Fe}{19}&  6.96&  15.1980&   15.205 (1.5)&     49.5 (7.5)&     54.5\\ 
\eli{Fe}{17}&  6.72&  15.2610&   15.266 (0.5)&    190.4 (10.8)&    174.6\\ 
\eli{Fe}{17}&  6.70&  15.4530&   15.459 (2.0)&     32.7 (6.9)&     28.8\\ 
\eli{Fe}{18}&  6.87&  15.4940&   15.493 (3.0)&     20.2 (6.4)&      9.4\\ 
\eli{Fe}{18}&  6.86&  15.6250&   15.629 (1.0)&     96.0 (9.2)&    118.6\\ 
\eli{Fe}{18}&  6.86&  15.8240&   15.829 (1.5)&     58.2 (8.5)&     71.0\\ 
\eli{Fe}{18}&  6.86&  15.8700&   15.876 (1.0)&     67.3 (8.9)&     42.8\\ 
\eli{Fe}{18}&  6.86&  16.0710&   16.079 (0.5)&    244.5 (14.2)&    199.9\\ 
\eli{Fe}{19}&  6.95&  16.1100&   16.112 (1.0)&     75.9 (9.8)&     92.6\\ 
\eli{Fe}{18}&  6.87&  16.1590&   16.169 (2.0)&     35.9 (8.0)&     52.1\\ 
\eli{Fe}{17}&  6.70&  16.7800&   16.781 (0.5)&    328.9 (17.7)&    351.5\\ 
\eli{Fe}{17}&  6.71&  17.0510&   17.056 (0.5)&    443.8 (21.3)&    413.6\\ 
\eli{Fe}{17}&  6.70&  17.0960&   17.101 (0.5)&    470.2 (21.8)&    476.8\\ 
\eli{Fe}{18}&  6.86&  17.6230&   17.627 (1.5)&    108.1 (13.9)&    121.7\\ 
\eli{O}{7}&  6.38&  17.7680&   17.758 (5.5)&     19.0 (9.8)&     13.2\\ 
\eli{O}{7}&  6.37&  18.6270&   18.633 (2.0)&     74.1 (14.1)&     37.3\\ 
\eli{Ca}{18}&  7.07&  18.6910&   18.685 (5.5)&     18.8 (11.5)&     25.5\\ 
\eli{O}{8}&  6.65&  18.9700&   18.974 (0.0)&   2142.0 (60.9)&   2140.0\\ 
\eli{Ca}{18}&  7.03&  19.6420&   19.627 (15.0)&      8.1 (10.8)&     18.2\\ 
\eli{Ca}{18}&  7.03&  19.7950&   19.793 (8.5)&     16.2 (13.5)&     36.3\\ 
\eli{N}{7}&  6.55&  19.8260&   19.830 (5.5)&     35.5 (15.8)&     29.6\\ 
\eli{N}{7}&  6.53&  20.9100&   20.912 (3.0)&    103.8 (23.7)&     99.6\\ 
\eli{O}{7}&  6.34&  21.6020&   21.605 (2.0)&    212.8 (32.5)&    247.0\\ 
\eli{O}{7}&  6.32&  21.8020&   21.815 (8.5)&     41.4 (22.4)&     35.3\\ 
\eli{O}{7}&  6.32&  22.0980&   22.101 (2.0)&    211.5 (37.1)&    155.1\\ 
\eli{N}{7}&  6.49&  24.7820&   24.785 (1.0)&    707.4 (62.0)&    667.3\\ 
  \enddata
  \tablecomments{\ Lines used in the EMD and abundance reconstruction
    for \siggem.}
  \tablenotetext{a}{$T_\mathrm{max}$ is the temperature of maximum
    emissivity according to AtomDB \citep{Smith:01,
      Foster:Smith:Brickhouse:2012}.}
  \tablenotetext{b}{The theoretical wavelength, from AtomDB.}
  \tablenotetext{c}{The
    measured wavelength; $1\sigma$ uncertainties are in
    parentheses in units of $\mmang$.}
  \tablenotetext{d}{The measured
    line flux; $1\sigma$ uncertainties are in parentheses.}
  \tablenotetext{e}{The theoretical line flux, from AtomDB, for the
    derived EMD and abundance model.}
\end{deluxetable}

\clearpage

\begin{deluxetable}{r c r r r r}
  \tablecolumns{6}
  \tablewidth{0.8\columnwidth}
  \tablecaption{\hrtennn Line Measurements\label{tbl:hrtenlines}}
  %
  \tablehead{
    \colhead{Ion}&
    \colhead{$\log T_\mathrm{max}$}&
    \colhead{$\lambda_0$}&
    \colhead{$\lambda_\mathrm{obs}$}&
    \colhead{$f$}&
    \colhead{$f_\mathrm{model}$}\\
    \colhead{}&
    \colhead{$\log\,$[K]}&
    \colhead{[$\rmang$]}&
    \colhead{[$\rmang$($\mathrm{m}\mang$)]}&
    \multicolumn{2}{c}{[$10^{-6}\apflux$]}\\
    \colhead{(1)}&
    \colhead{(2)}&
    \colhead{(3)}&
    \colhead{(4)}&
    \colhead{(5)}&
    \colhead{(6)}
  }
  \startdata
\eli{Fe}{25}&  7.82&   1.8607&    1.858 (1.4)&     83.4 (8.0)&     68.5\\ 
\eli{Ca}{19}&  7.50&   3.1772&    3.176 (2.0)&     16.8 (3.2)&     13.5\\ 
\eli{Ca}{19}&  7.46&   3.1909&    3.192 (3.6)&      9.8 (2.9)&      4.4\\ 
\eli{Ca}{19}&  7.46&   3.2110&    3.210 (3.4)&      7.4 (2.5)&      4.5\\ 
\eli{K}{18}&  7.42&   3.5273&    3.521 (5.2)&      6.8 (2.5)&      0.8\\ 
\eli{K}{18}&  7.36&   3.5434&    3.545 (0.0)&      2.3 (2.3)&      0.1\\ 
\eli{K}{18}&  7.39&   3.5669&    3.568 (0.0)&      6.0 (2.5)&      0.4\\ 
\eli{Ar}{18}&  7.73&   3.7338&    3.736 (0.7)&     28.4 (2.9)&     29.5\\ 
\eli{Ar}{17}&  7.36&   3.9491&    3.950 (0.6)&     38.7 (3.4)&     36.8\\ 
\eli{Ar}{17}&  7.31&   3.9676&    3.968 (1.1)&     17.8 (2.9)&     10.0\\ 
\eli{S}{16}&  7.51&   3.9923&    3.995 (0.9)&     25.1 (3.0)&     11.8\\ 
\eli{S}{16}&  7.57&   4.7301&    4.731 (0.6)&     47.3 (3.8)&     52.4\\ 
\eli{Si}{14}&  7.43&   4.9462&    4.948 (1.9)&      8.4 (2.9)&      4.3\\ 
\eli{S}{15}&  7.20&   5.0387&    5.041 (0.5)&     58.4 (4.2)&     56.3\\ 
\eli{S}{15}&  7.16&   5.0648&    5.068 (2.0)&     13.2 (3.2)&     12.8\\ 
\eli{S}{15}&  7.17&   5.1015&    5.101 (1.0)&     31.3 (3.7)&     24.7\\ 
\eli{Si}{14}&  7.42&   5.2174&    5.220 (0.9)&     26.2 (3.8)&     24.1\\ 
\eli{Si}{13}&  7.07&   5.6805&    5.685 (1.3)&     17.1 (3.1)&     16.9\\ 
\eli{Si}{14}&  7.40&   6.1831&    6.185 (0.2)&    162.5 (3.9)&    164.8\\ 
\eli{Si}{13}&  7.03&   6.6479&    6.651 (0.2)&    125.5 (3.5)&    120.2\\ 
\eli{Si}{13}&  6.99&   6.6866&    6.690 (0.6)&     26.8 (2.4)&     22.9\\ 
\eli{Si}{13}&  7.00&   6.7403&    6.742 (0.3)&     82.0 (2.9)&     58.2\\ 
\eli{Mg}{12}&  7.22&   7.1063&    7.109 (0.6)&     26.7 (2.3)&     30.6\\ 
\eli{Al}{13}&  7.38&   7.1714&    7.172 (0.7)&     30.0 (2.4)&     29.0\\ 
\eli{Al}{12}&  6.94&   7.7573&    7.761 (1.1)&     16.1 (2.0)&      8.8\\ 
\eli{Mg}{11}&  6.87&   7.8503&    7.853 (1.3)&     12.3 (2.1)&     11.5\\ 
\eli{Fe}{24}&  7.46&   7.9857&    7.986 (0.9)&     22.2 (2.4)&     19.0\\ 
\eli{Fe}{24}&  7.46&   7.9960&    7.998 (1.1)&     12.3 (2.4)&      9.5\\ 
\eli{Fe}{23}&  7.28&   8.3038&    8.307 (1.3)&     17.8 (3.2)&     19.4\\ 
\eli{Fe}{24}&  7.44&   8.3161&    8.320 (0.9)&     26.4 (3.4)&     19.1\\ 
\eli{Fe}{24}&  7.44&   8.3761&    8.378 (2.0)&      9.0 (2.5)&      7.7\\ 
\eli{Mg}{12}&  7.19&   8.4219&    8.424 (0.2)&    202.8 (5.3)&    208.2\\ 
\eli{Fe}{21}&  7.10&   8.5740&    8.578 (1.2)&     13.3 (2.3)&     11.2\\ 
\eli{Fe}{23}&  7.27&   8.8149&    8.819 (0.8)&     24.0 (2.5)&     22.0\\ 
\eli{Fe}{22}&  7.17&   8.9748&    8.979 (0.8)&     21.2 (2.4)&     20.8\\ 
\eli{Mg}{11}&  6.84&   9.1687&    9.172 (0.3)&    112.1 (4.1)&     79.0\\ 
\eli{Fe}{21}&  7.10&   9.1944&    9.192 (1.9)&     20.4 (2.9)&      8.4\\ 
\eli{Mg}{11}&  6.80&   9.2297&    9.233 (1.3)&     19.7 (2.8)&     12.7\\ 
\eli{Mg}{11}&  6.81&   9.3143&    9.318 (0.4)&     56.5 (3.1)&     39.5\\ 
\eli{Ne}{10}&  6.99&   9.4808&    9.481 (0.4)&     64.0 (2.7)&     38.9\\ 
\eli{Fe}{19}&  6.97&   9.6951&    9.698 (2.7)&     18.1 (3.8)&      7.6\\ 
\eli{Ne}{10}&  6.98&   9.7082&    9.712 (0.7)&     90.5 (4.8)&     89.7\\ 
\eli{Na}{11}&  7.08&  10.0240&   10.031 (1.0)&     26.9 (3.1)&     23.2\\ 
\eli{Ni}{19}&  6.87&  10.1100&   10.112 (4.0)&      9.7 (4.7)&      3.9\\ 
\eli{Ne}{10}&  6.97&  10.2390&   10.242 (0.0)&    260.3 (6.2)&    252.8\\ 
\eli{Fe}{24}&  7.45&  10.6190&   10.625 (0.5)&    145.8 (5.4)&    134.3\\ 
\eli{Fe}{24}&  7.45&  10.6630&   10.665 (0.5)&     69.9 (4.2)&     69.4\\ 
\eli{Fe}{19}&  6.97&  10.8160&   10.823 (1.0)&     29.3 (3.3)&     27.1\\ 
\eli{Fe}{23}&  7.27&  10.9810&   10.985 (0.0)&     93.1 (4.8)&    104.8\\ 
\eli{Ne}{9}&  6.66&  11.0010&   11.005 (1.5)&     35.6 (4.2)&     22.1\\ 
\eli{Fe}{23}&  7.27&  11.0190&   11.024 (1.0)&     76.3 (7.2)&     67.2\\ 
\eli{Fe}{24}&  7.42&  11.0290&   11.036 (0.5)&     72.9 (7.3)&     88.3\\ 
\eli{Fe}{17}&  6.76&  11.1310&   11.138 (2.0)&     18.9 (3.8)&     19.8\\ 
\eli{Fe}{24}&  7.42&  11.1760&   11.179 (0.0)&    166.7 (6.7)&    159.2\\ 
\eli{Fe}{17}&  6.76&  11.2540&   11.255 (1.5)&     24.6 (4.9)&     27.4\\ 
\eli{Fe}{24}&  7.42&  11.2680&   11.268 (1.0)&     46.9 (5.2)&     36.5\\ 
\eli{Fe}{18}&  6.89&  11.3260&   11.330 (0.5)&     48.1 (4.0)&     38.7\\ 
\eli{Fe}{18}&  6.88&  11.5270&   11.531 (1.0)&     48.9 (4.4)&     37.8\\ 
\eli{Ne}{9}&  6.64&  11.5440&   11.551 (0.5)&     68.7 (4.8)&     63.6\\ 
\eli{Fe}{23}&  7.26&  11.7360&   11.744 (0.5)&    238.8 (7.4)&    223.9\\ 
\eli{Fe}{22}&  7.16&  11.7700&   11.775 (0.0)&    209.9 (7.0)&    188.6\\ 
\eli{Fe}{22}&  7.15&  11.9320&   11.937 (1.0)&     65.7 (4.9)&     27.2\\ 
\eli{Ne}{10}&  6.94&  12.1350&   12.137 (3.5)&   1839.0 (19.3)&   1838.2\\ 
\eli{Fe}{23}&  7.25&  12.1610&   12.162 (1.0)&    120.4 (9.2)&    119.6\\ 
\eli{Fe}{17}&  6.75&  12.2660&   12.268 (1.0)&     70.9 (9.7)&     60.0\\ 
\eli{Fe}{21}&  7.09&  12.2840&   12.289 (0.5)&    304.5 (12.2)&    342.4\\ 
\eli{Fe}{22}&  7.15&  12.7540&   12.756 (0.5)&     79.9 (6.6)&     64.5\\ 
\eli{Fe}{20}&  7.03&  12.8240&   12.831 (1.0)&    140.1 (10.0)&     60.1\\ 
\eli{Fe}{20}&  7.03&  12.8460&   12.849 (1.0)&    123.9 (14.1)&    140.4\\ 
\eli{Fe}{20}&  7.03&  13.3850&   13.381 (2.0)&     47.1 (9.2)&     33.0\\ 
\eli{Fe}{19}&  6.96&  13.4230&   13.434 (2.5)&     42.2 (7.9)&     24.1\\ 
\eli{Ne}{9}&  6.61&  13.4470&   13.452 (0.5)&    414.6 (14.3)&    443.3\\ 
\eli{Fe}{19}&  6.96&  13.4620&   13.471 (1.0)&     92.5 (8.7)&     55.8\\ 
\eli{Fe}{19}&  6.96&  13.5180&   13.527 (0.5)&    188.4 (12.1)&    211.4\\ 
\eli{Ne}{9}&  6.58&  13.5520&   13.558 (1.0)&    101.5 (9.2)&     71.7\\ 
\eli{Fe}{19}&  6.96&  13.6450&   13.654 (1.5)&     55.9 (6.9)&     33.8\\ 
\eli{Ne}{9}&  6.59&  13.6990&   13.703 (0.5)&    279.7 (11.5)&    249.3\\ 
\eli{Fe}{20}&  7.02&  13.7670&   13.771 (2.5)&     54.8 (7.3)&     23.2\\ 
\eli{Fe}{19}&  6.96&  13.7950&   13.800 (1.0)&    102.6 (8.6)&     90.1\\ 
\eli{Fe}{18}&  6.87&  14.2080&   14.209 (0.5)&    309.1 (14.3)&    387.5\\ 
\eli{Fe}{18}&  6.87&  14.2560&   14.263 (1.0)&     87.1 (7.1)&     76.7\\ 
\eli{Fe}{20}&  7.02&  14.2670&   14.276 (1.5)&     54.0 (12.1)&     39.3\\ 
\eli{Fe}{18}&  6.87&  14.3430&   14.350 (1.5)&     52.1 (7.9)&     43.0\\ 
\eli{Fe}{18}&  6.87&  14.3730&   14.379 (1.0)&    112.8 (10.0)&     92.5\\ 
\eli{Fe}{18}&  6.87&  14.5340&   14.541 (1.0)&    115.8 (12.8)&     76.8\\ 
\eli{Fe}{19}&  6.96&  14.6640&   14.671 (1.0)&     72.4 (9.6)&     85.0\\ 
\eli{O}{8}&  6.71&  14.8210&   14.823 (1.0)&     60.4 (8.5)&     46.9\\ 
\eli{Fe}{17}&  6.73&  15.0140&   15.017 (0.5)&    542.6 (18.0)&    563.4\\ 
\eli{Fe}{19}&  6.95&  15.0790&   15.085 (1.0)&    118.9 (9.5)&     87.4\\ 
\eli{O}{8}&  6.70&  15.1760&   15.180 (1.0)&    118.1 (9.3)&    106.7\\ 
\eli{Fe}{19}&  6.96&  15.1980&   15.205 (1.5)&     49.5 (7.5)&     54.5\\ 
\eli{Fe}{17}&  6.72&  15.2610&   15.266 (0.5)&    190.4 (10.8)&    174.6\\ 
\eli{Fe}{17}&  6.70&  15.4530&   15.459 (2.0)&     32.7 (6.9)&     28.8\\ 
\eli{Fe}{18}&  6.87&  15.4940&   15.493 (3.0)&     20.2 (6.4)&      9.4\\ 
\eli{Fe}{18}&  6.86&  15.6250&   15.629 (1.0)&     96.0 (9.2)&    118.6\\ 
\eli{Fe}{18}&  6.86&  15.8240&   15.829 (1.5)&     58.2 (8.5)&     71.0\\ 
\eli{Fe}{18}&  6.86&  15.8700&   15.876 (1.0)&     67.3 (8.9)&     42.8\\ 
\eli{Fe}{18}&  6.86&  16.0710&   16.079 (0.5)&    244.5 (14.2)&    199.9\\ 
\eli{Fe}{19}&  6.95&  16.1100&   16.112 (1.0)&     75.9 (9.8)&     92.6\\ 
\eli{Fe}{18}&  6.87&  16.1590&   16.169 (2.0)&     35.9 (8.0)&     52.1\\ 
\eli{Fe}{17}&  6.70&  16.7800&   16.781 (0.5)&    328.9 (17.7)&    351.5\\ 
\eli{Fe}{17}&  6.71&  17.0510&   17.056 (0.5)&    443.8 (21.3)&    413.6\\ 
\eli{Fe}{17}&  6.70&  17.0960&   17.101 (0.5)&    470.2 (21.8)&    476.8\\ 
\eli{Fe}{18}&  6.86&  17.6230&   17.627 (1.5)&    108.1 (13.9)&    121.7\\ 
\eli{O}{7}&  6.38&  17.7680&   17.758 (5.5)&     19.0 (9.8)&     13.2\\ 
\eli{O}{7}&  6.37&  18.6270&   18.633 (2.0)&     74.1 (14.1)&     37.3\\ 
\eli{Ca}{18}&  7.07&  18.6910&   18.685 (5.5)&     18.8 (11.5)&     25.5\\ 
\eli{O}{8}&  6.65&  18.9700&   18.974 (0.0)&   2142.0 (60.9)&   2140.0\\ 
\eli{Ca}{18}&  7.03&  19.6420&   19.627 (15.0)&      8.1 (10.8)&     18.2\\ 
\eli{Ca}{18}&  7.03&  19.7950&   19.793 (8.5)&     16.2 (13.5)&     36.3\\ 
\eli{N}{7}&  6.55&  19.8260&   19.830 (5.5)&     35.5 (15.8)&     29.6\\ 
\eli{N}{7}&  6.53&  20.9100&   20.912 (3.0)&    103.8 (23.7)&     99.6\\ 
\eli{O}{7}&  6.34&  21.6020&   21.605 (2.0)&    212.8 (32.5)&    247.0\\ 
\eli{O}{7}&  6.32&  21.8020&   21.815 (8.5)&     41.4 (22.4)&     35.3\\ 
\eli{O}{7}&  6.32&  22.0980&   22.101 (2.0)&    211.5 (37.1)&    155.1\\ 
\eli{N}{7}&  6.49&  24.7820&   24.785 (1.0)&    707.4 (62.0)&    667.3\\ 
  \enddata
 \tablecomments{\ Lines used in the EMD and abundance reconstruction for
   \hrtennn.  Columns are analogous to those in Table~\ref{tbl:sgemlines}.}
\end{deluxetable}

\clearpage

\appendix
\section{Supplemental Material}

\subsection{Line Flux Ratio Residuals}\label{app:ratioresids}

\begin{figure}[!htb]
  \centering\leavevmode
  \includegraphics[width=0.45\columnwidth, viewport=40 20 500 460]{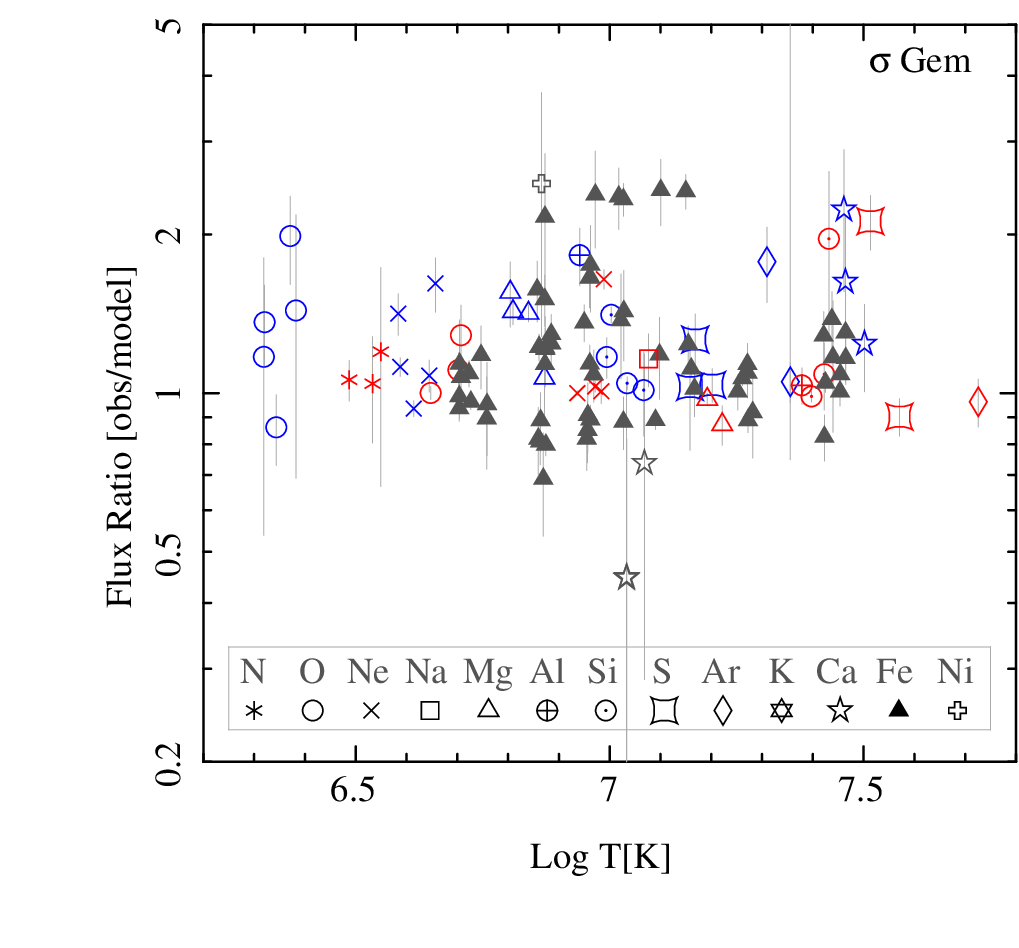}
  \includegraphics[width=0.45\columnwidth, viewport=40 20 500 460]{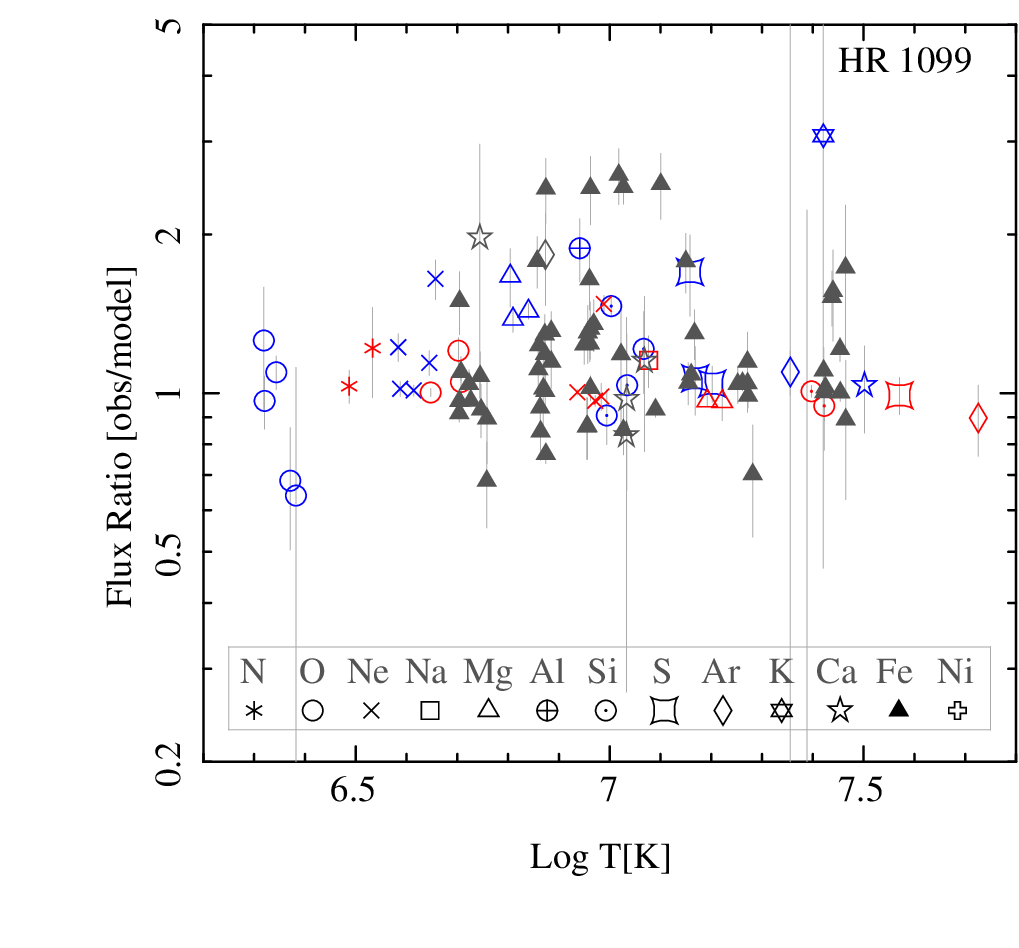}\newline

  \includegraphics[width=0.45\columnwidth, viewport=40 20 500 460]{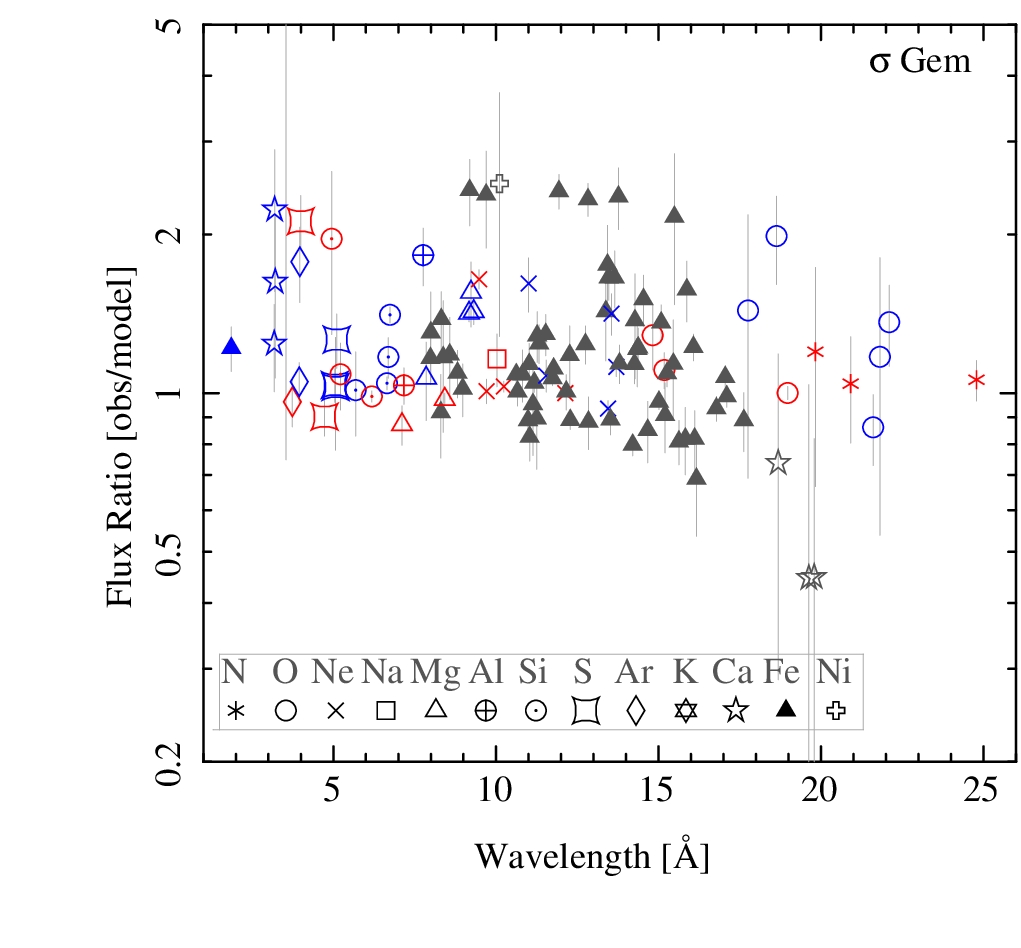}
  \includegraphics[width=0.45\columnwidth, viewport=40 20 500 460]{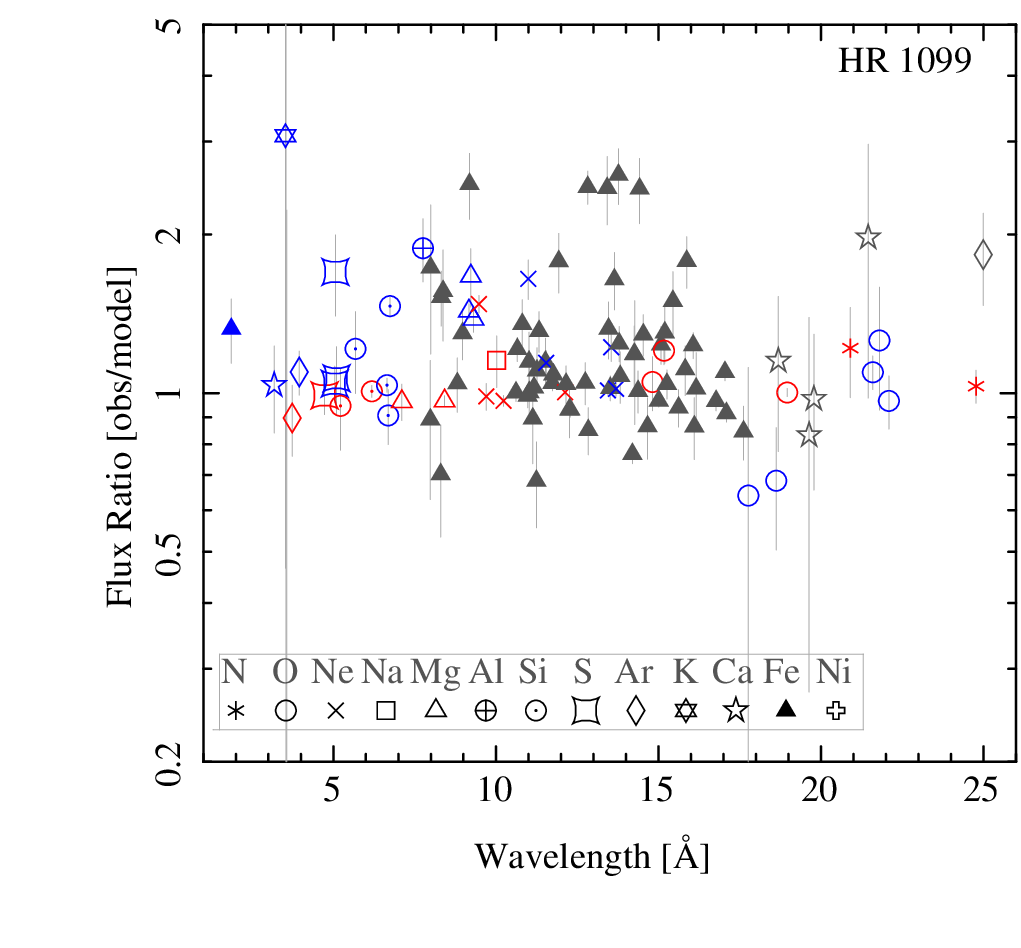}\newline

  \includegraphics[width=0.45\columnwidth, viewport=40 20 500 460]{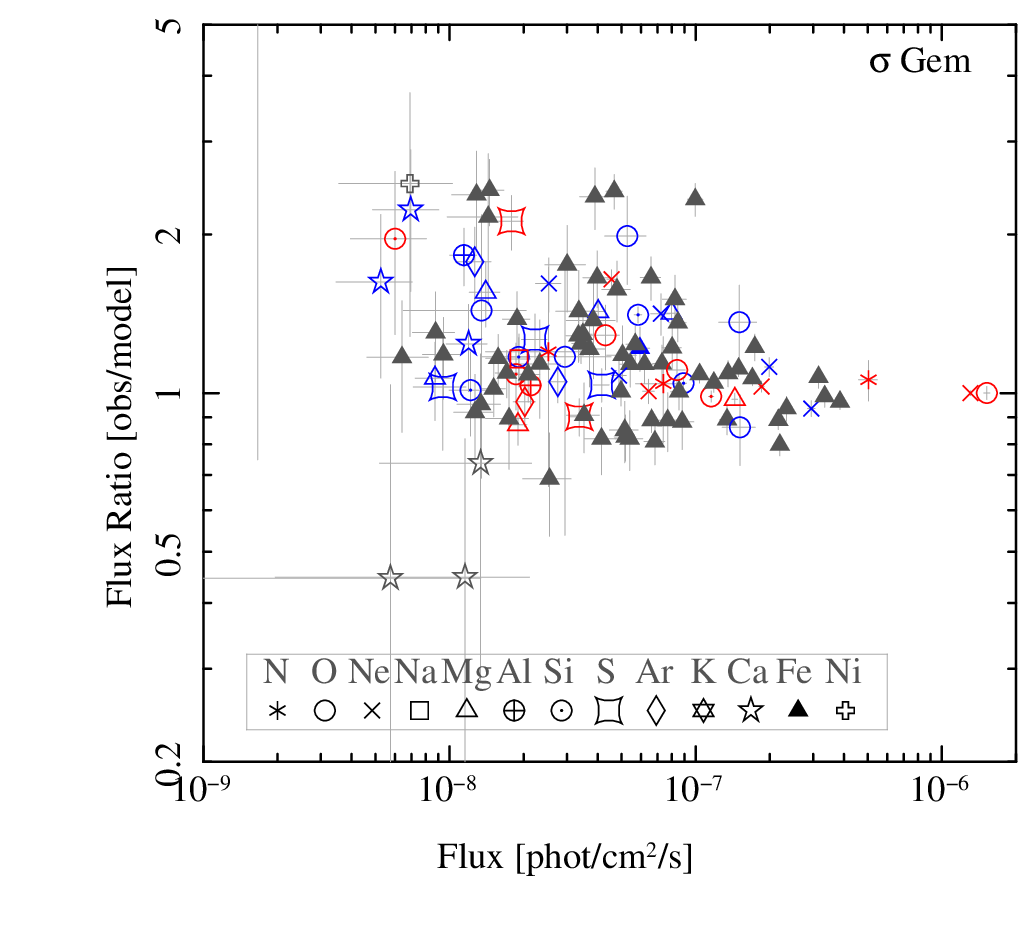}
  \includegraphics[width=0.47\columnwidth, viewport=40 20 500 460]{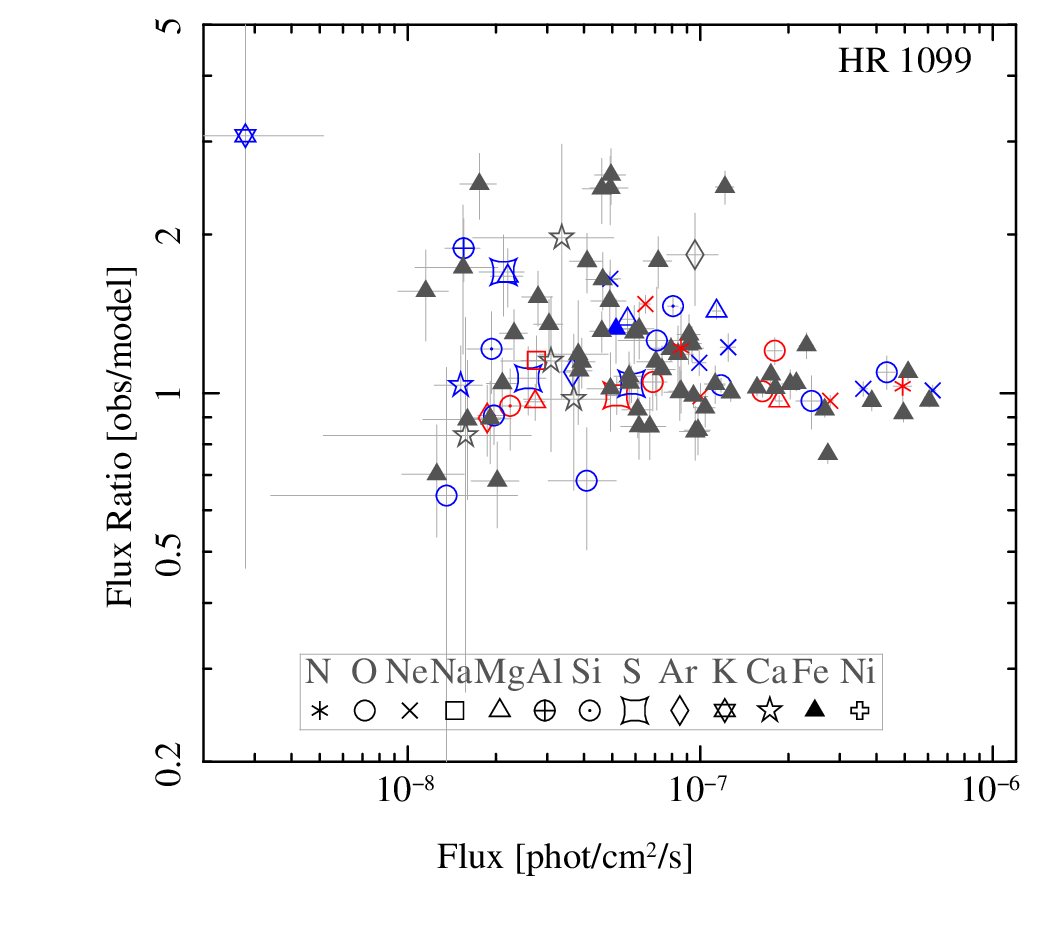}\newline

  \caption{Line flux ratio residuals for \siggem (left column) and
    \hrten (right column) against temperature of peak emissivity (top
    row), wavelength (middle row), and line flux (bottom row).
    Elements are plotted with different symbols.  H-like lines are in
    red, He-like are blue, and others (primarily Fe) are gray.}
  \label{fig:lineratioresids}

\end{figure}

Figure~\ref{fig:lineratioresids} shows in detail the line flux ratio
residuals (model to data) against wavelength, temperature, and line
flux.  Features with ratios roughly between 0.5 and 2.0 were used in
the emission measure modeling.  We include some of the weaker lines
(such as of K, Na, and Al) which were not used, but were later fit for
abundances post facto using the emission measure solution.  The
figures show that there are a significant number of lines spanning the
wavelength and temperature ranges, and of significant quality for
emission measure modeling.

\begin{figure}[!htb]
  \centering\leavevmode
  \includegraphics[width=0.45\columnwidth]{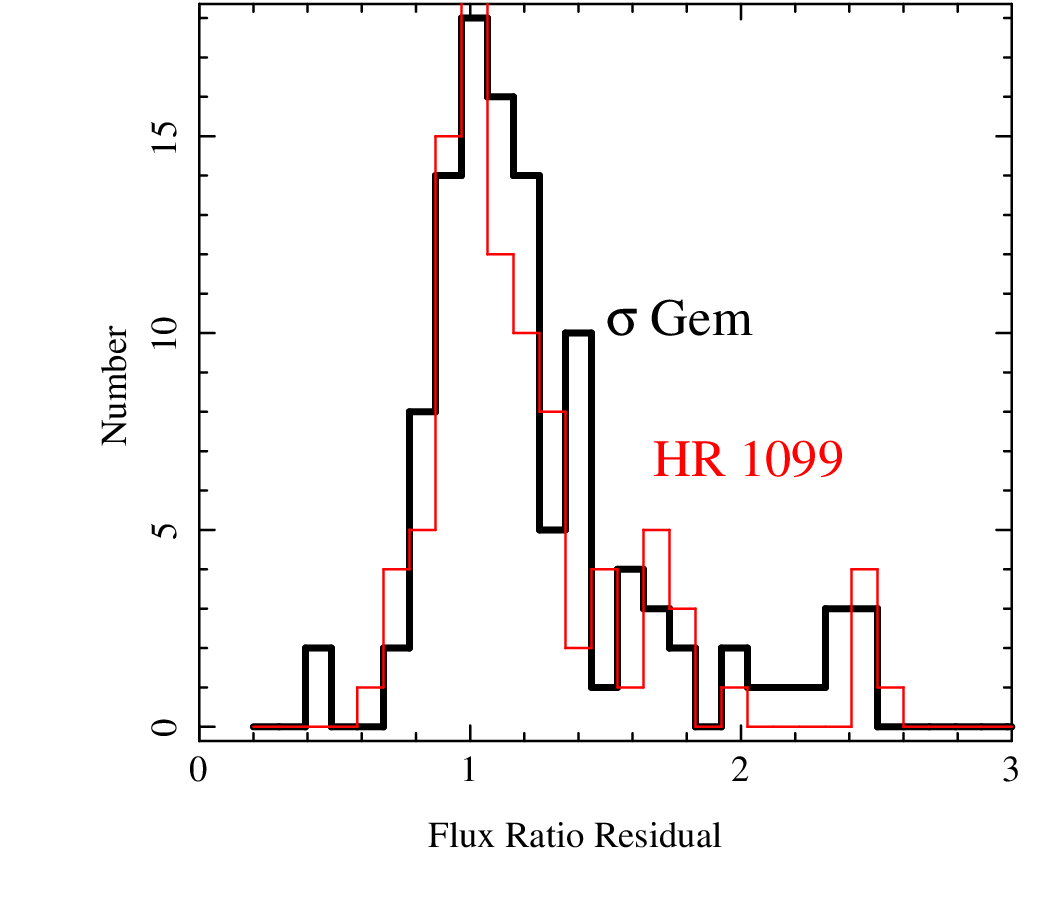}

  \caption{Histogram of the line flux ratio residuals for \siggem
    (thicker or dark line) and
    \hrten (thinner or red line).  There is a small systematic bias
    for outliers to have high residuals.  This is likely due to
    unresolved blends in weak lines making the measured flux higher
    than expected.}
  \label{fig:residhist}

\end{figure}

{   
A small bias can be seen in that residuals are slightly more likely to
be high than low.  This is more clearly seen in the histogram of the
residuals in Figure~\ref{fig:residhist}.  There is a tail above a
ratio of 1.4, and it is similar in both stars.  Looking at the lower
panel of Figure~\ref{fig:lineratioresids} we can see that this bias is
more prevalent in the weaker lines.  Hence, we believe it is due to
inclusion of unresolved blends in the measured flux which are not
accounted for in the model flux.  A systematically low continuum would
also produce high residuals, but the flat residuals seen in
Figures~\ref{fig:threespec}, \ref{fig:kregdetail},
\ref{fig:naregdetail} and~\ref{fig:siggemspec} argue against that
explanation, as do other well modeled weak lines.
} 

\clearpage 

\subsection{Emission Measure Tables}

The following tables give the emission measure distributions for
\siggem (Table~\ref{tbl:emdsiggem}), \hrten
(Table~\ref{tbl:emdhrten}), and the solar flare
(Table~\ref{tbl:emdsun}).

\begin{deluxetable}{c r r r}
  \tablecolumns{4}
  \tablewidth{0.8\columnwidth}
  \tablecaption{\siggem Emission Measure Distribution\label{tbl:emdsiggem}}
  \tablehead{
    \colhead{$\log T$}&
    \colhead{$EM_{51}$}&
    \colhead{$EM_{low}$}&
    \colhead{$EM_{high}$}\\
    \colhead{[$\log\,$K]}&
    \multicolumn{3}{c}{[$10^{51}\cmmthree$]}\\
    \colhead{(1)}&
    \colhead{(2)}&
    \colhead{(3)}&
    \colhead{(4)}
  }
  \startdata
   6.3&       2.4&      1.7&     3.2\\
   6.4&       8.5&      6.2&    10.9\\
   6.5&      21.9&     18.8&    25.0\\
   6.6&      36.9&     32.8&    41.0\\
   6.7&      35.4&     31.5&    39.3\\
   6.8&      61.1&     55.0&    67.2\\
   6.9&      84.5&     77.2&    91.8\\
   7.0&     304.5&    292.2&   316.7\\
   7.1&     248.2&    231.5&   264.9\\
   7.2&     392.1&    370.3&   413.9\\
   7.3&     301.3&    279.8&   322.8\\
   7.4&     138.7&    124.4&   153.0\\
   7.5&     109.0&     91.6&   126.4\\
   7.6&     157.6&    134.6&   180.6\\
   7.7&     240.2&    211.0&   269.5\\
   7.8&     182.5&    136.9&   228.1\\
   7.9&      60.1&     41.4&    78.8\\
   8.0&      13.6&      9.6&    17.5\\
   8.1&       3.3&      2.4&     4.1\\
   8.2&       1.0&      0.8&     1.2\\
   8.3&       0.5&      0.4&     0.6
  \enddata
 \tablecomments{\ Emission measure values corresponding to the data
   plotted in Figure~\ref{fig:emd}.  Values are integrated over
   uniform logarithmic temperature bins of
   $0.1\,\mathrm{dex}$. Columns 3 and 4 give the $1\sigma$ statistical
   uncertainties based on line-flux uncertainties and Monte-Carlo
   emission measure reconstruction runs.}
\end{deluxetable}

\begin{deluxetable}{c r r r}
  \tablecolumns{4}
  \tablewidth{0.8\columnwidth}
  \tablecaption{\hrten Emission Measure Distribution\label{tbl:emdhrten}}
  \tablehead{
    \colhead{$\log T$}&
    \colhead{$EM_{51}$}&
    \colhead{$EM_{low}$}&
    \colhead{$EM_{high}$}\\
    \colhead{[$\log\,$K]}&
    \multicolumn{3}{c}{[$10^{51}\cmmthree$]}\\
    \colhead{(1)}&
    \colhead{(2)}&
    \colhead{(3)}&
    \colhead{(4)}
  }
  \startdata
  6.3&         2.1&        1.1&        2.1\\
  6.4&         8.5&        4.1&        7.3\\
  6.5&        25.9&       12.6&       16.8\\
  6.6&        39.4&       22.0&       27.4\\
  6.7&        34.5&       21.1&       26.3\\
  6.8&        41.1&       36.8&       45.0\\
  6.9&        94.4&       51.7&       61.5\\
  7.0&       179.9&      195.6&      212.0\\
  7.1&       169.3&      154.9&      177.3\\
  7.2&       198.3&      247.9&      277.1\\
  7.3&       233.5&      187.3&      216.1\\
  7.4&       205.3&       83.2&      102.4\\
  7.5&       151.5&       61.3&       84.6\\
  7.6&       103.4&       90.1&      120.9\\
  7.7&        66.4&      141.2&      180.4\\
  7.8&        38.1&       91.7&      152.7\\
  7.9&        19.5&       27.7&       52.8\\
  8.0&         8.9&        6.4&       11.7\\
  8.1&         3.9&        1.6&        2.7\\
  8.2&         1.7&        0.5&        0.8\\
  8.3&         0.8&        0.3&        0.4
  \enddata
 \tablecomments{\ Emission measure values corresponding to the data
   plotted in Figure~\ref{fig:emd}.  Values are integrated over
   uniform logarithmic temperature bins of
   $0.1\,\mathrm{dex}$. Columns 3 and 4 give the $1\sigma$ statistical
   uncertainties based on line-flux uncertainties and Monte-Carlo
   emission measure reconstruction runs.}
\end{deluxetable}

\clearpage
\begin{deluxetable}{c c r}
  \tablecolumns{3}
  \tablewidth{0.8\columnwidth}
  \tablecaption{Solar Flare Emission Measure\label{tbl:emdsun}}
  \tablehead{
    \colhead{$\log T_{low}$}&
    \colhead{$\log T_{high}$}&
    \colhead{$EM_{57}$}\\
    \multicolumn{2}{c}{[$\log\,$K]}&
    \colhead{[$10^{57}\cmmthree$]}\\
    \colhead{(1)}&
    \colhead{(2)}&
    \colhead{(3)}
  }
  \startdata
  6.50&   6.53&      19170.53\\
  6.53&   6.57&       8624.91\\
  6.57&   6.60&        956.94\\
  6.60&   6.64&         24.58\\
  6.64&   6.67&          0.17\\
  6.67&   6.92&          0.00\\
  6.92&   7.17&          0.03\\
  7.17&   7.21&          0.00\\
  7.21&   7.24&          0.29\\
  7.24&   7.28&         59.82\\
  7.28&   7.32&        467.73\\
  7.32&   7.35&         97.59\\
  7.35&   7.39&          0.60\\
  7.39&   7.42&          0.01
  \enddata
  \tablecomments{\ Emission measure values corresponding to the data
    plotted in Figure~\ref{fig:emd}.  Values are integrated over
    variable-width logarithmic temperature bins.}
\end{deluxetable}


\end{document}